\documentclass[fleqn,usenatbib]{mnras}

\usepackage{newtxtext,newtxmath}
\usepackage[T1]{fontenc}

\DeclareRobustCommand{\VAN}[3]{#2}
\let\VANthebibliography\thebibliography
\def\thebibliography{\DeclareRobustCommand{\VAN}[3]{##3}\VANthebibliography}

\usepackage{float, graphicx}	% Including figure files
\graphicspath{{./}{Figures/}}
\usepackage{amsmath}	% Advanced maths commands
\usepackage{enumitem}
\newlist{todolist}{itemize}{2}
\setlist[todolist]{label=$\square$}
\usepackage{xcolor}
\usepackage{array}
\usepackage{orcidlink}

\title[Observational Signatures of C/O WD Merger Remnants]{Observational Signatures of Carbon-Oxygen White Dwarf Merger Remnants}

\author[Yao et al.]{
Philippe Z. Yao\,\orcidlink{0000-0003-3024-7218},$^{1}$\thanks{E-mail: philippe.yao@princeton.edu}
Eliot Quataert\,\orcidlink{0000-0001-9185-5044},$^{1}$
Andy Goulding\,\orcidlink{0000-0003-4700-663X}$^{1}$
\\
$^{1}$Department of Astrophysical Sciences, Princeton University, Peyton Hall, Princeton, NJ 08544, USA}

\date{Accepted XXX. Received YYY; in original form ZZZ}

\pubyear{2015}

\begin{document}
\label{firstpage}
\pagerange{\pageref{firstpage}--\pageref{lastpage}}
\maketitle

% Abstract of the paper
\begin{abstract}
    Many double white dwarf (WD) mergers likely do not lead to a prompt thermonuclear explosion.   We investigate the prospects for observationally detecting the surviving remnants of such mergers, focusing on the case of mergers of double Carbon-Oxygen WDs.   For $\sim 10^4$ yr, the merger remnant is observationally similar to an extreme AGB star evolving to become a massive WD.  Identifying merger remnants is thus easiest in galaxies with high stellar masses (high WD merger rate) and low star formation rates (low birth rate of $\sim 6-10 \,{\rm M_{\odot}}$ stars). Photometrically identifying merger remnants is challenging even in these cases because the merger remnants appear similar to He stars and post-outburst classical novae. We propose that the most promising technique for discovering WD merger remnants is through their unusual surrounding photoionized nebulae. We use CLOUDY photoionization calculations to investigate their unique spectral features. Merger remnants should produce weak hydrogen lines and strong carbon and oxygen recombination and fine-structure lines in the UV, optical and IR.  With narrow-band imaging or integral field spectrographs, we predict that multiple candidates are detectable in the bulge of M31, the outskirts of M87 and other nearby massive galaxies, and the Milky Way. Our models roughly reproduce the WISE nebula surrounding the Galactic WD merger candidate IRAS 00500+6713; we predict detectable [Ne\,VI] and [Mg\,VII] lines with JWST but that the mid-IR WISE emission is dominated by dust not fine-structure lines.
\end{abstract}

% Select between one and six entries from the list of approved keywords.
% Don't make up new ones.
\begin{keywords}
white dwarfs -- supernovae: general -- planetary nebulae: general -- stars: evolution
\end{keywords}

%%%%%%%%%%%%%%%%%%%%%%%%%%%%%%%%%%%%%%%%%%%%%%%%%%

%%%%%%%%%%%%%%%%% BODY OF PAPER %%%%%%%%%%%%%%%%%%

\section{Introduction}

The merger of two white dwarfs (WDs) can result in a wide range of outcomes depending on  the total mass, mass ratio, and chemical composition of the system.   Although some double WD mergers likely lead to a thermonuclear explosion and a Type Ia supernova \citep{Webbink1984,Iben1984}, many probably do not and instead leave behind a long-lived remnant. If the total mass of the merger remnant remains above the Chandrasekhar mass, the merger remnant ultimately develops an iron core and collapses to form a neutron star \citep{Schwab2016}.   Lower mass merger remnants can evolve to become a cooling WD, an R Coronae Borealis star, or a hot subdwarf \citep[][respectively]{Hollands2020,Staff2012,Schwab2018}. In this paper, we will focus on the merger of two C/O WDs or a C/O and O/Ne/Mg WD, which will not result in R Coronae Borealis stars or hot subdwarfs.  In WD merger models, the former is thought to originate from He+C/O WD mergers \citep[e.g.][]{Webbink1984, Saio2002,Munson2021}, and the latter from double He WD mergers \citep[e.g.][]{Webbink1984, Iben1986}.

\begin{figure*}
        \centering
        \includegraphics[width=\textwidth]{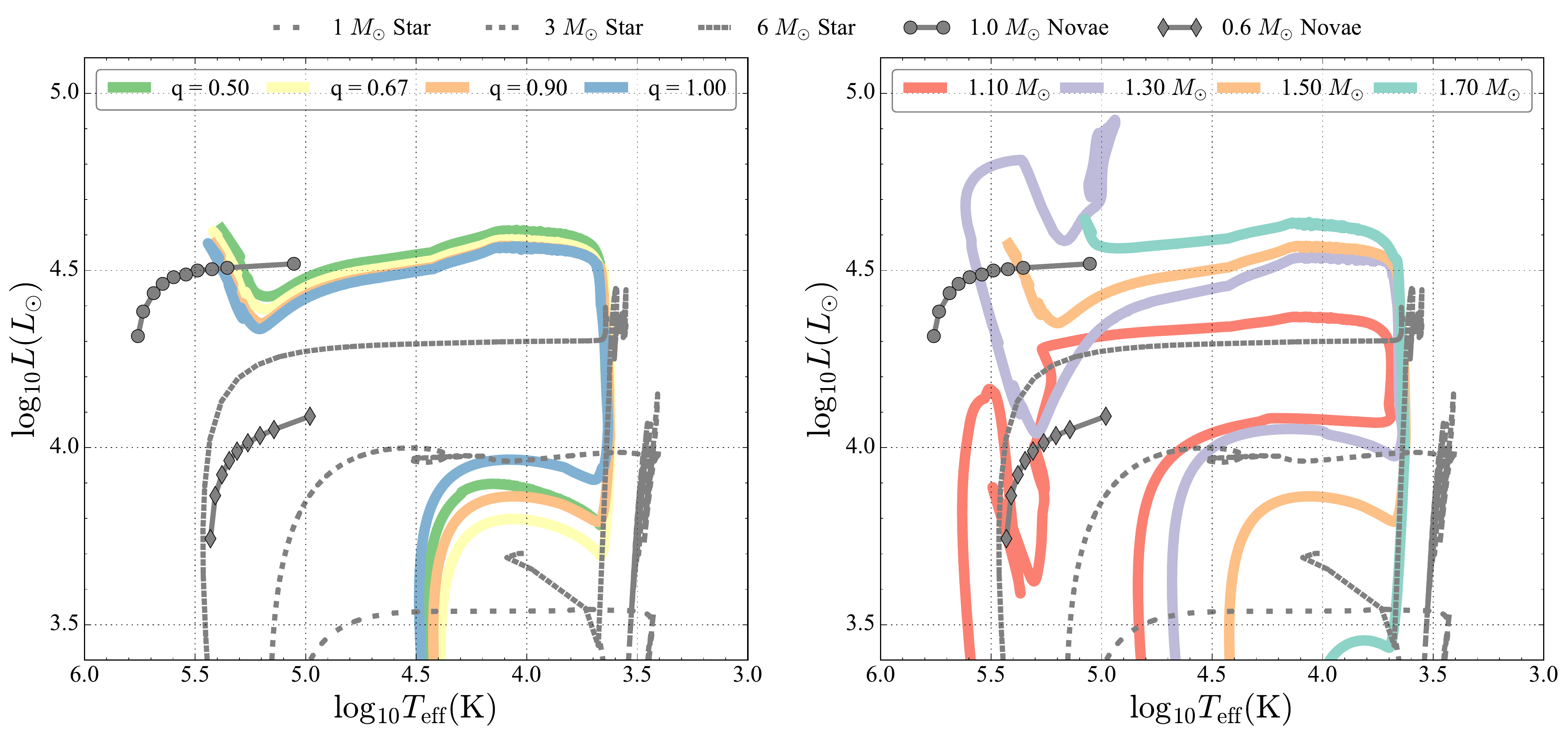}
        \caption{C/O WD merger remnants and potential stellar contaminants in the H-R diagram. The left panel shows models from \citet{Schwab2021} with fixed total mass ($1.5 {\rm M_{\odot}}$) and different mass ratios q. The right panel shows models from \citet{Schwab2021} with fixed mass ratios (q = 0.9) and different total masses. The orange line is the same in both panels. MESA stellar tracks of typical post-main-sequence stars evolving into cooling WDs with a range of masses from \citet{Farag2020} are plotted with dotted lines. The grey solid line shows the trajectory of post-outburst classical novae from \citet{Wolf2013}; markers are separated by equal time portions of 1 year and 30 years for $1 {\rm M_{\odot}}$ and $0.6 {\rm M_{\odot}}$ progenitors respectively.  Post-outburst novae and $\sim 6-10{\rm M_{\odot}}$ post-AGB stars are similar in $L-T_{\rm eff}$ space to WD merger remnants.}
        \label{fig:contaminants}
    \end{figure*}

The timescale for the merger remnant to evolve towards its final fate is $\sim 10^{3-4}$ yr, set by the thermal time needed to radiate away the excess thermal energy created during the merger \citep{Shen2012}.  During this thermal timescale evolution, the remnant initially evolves to become a giant and then contracts and heats up to become a hot proto-WD \citep{Shen2012,Schwab2016,Schwab2021}.   The resulting evolution is similar to that of a $\sim 6-10{\rm M_{\odot}}$ star as it evolves from the AGB to become a cooling WD.   To highlight this similarity, Figure \ref{fig:contaminants} shows the post-merger evolution of double WD merger remnants predicted by \citet{Schwab2021} compared to the evolution of $1, 3$, and $6{\rm M_{\odot}}$ stars (we also show post outburst classical novae for reasons discussed later in the paper).  There is considerable overlap in the luminosity and effective temperatures of the WD merger remnants and the 6 $M_\odot$ star (the similarity would be even stronger for stars with $7-10{\rm M_{\odot}}$, but those models are very difficult to evolve to late stages because of off-center nuclear burning; \citealt{Jones2013}).

As shown in Figure \ref{fig:contaminants}, theoretical models predict that surviving WD merger remnants should evolve through a giant phase to become a hot, bright proto-WD \citep{Shen2012,Schwab2016,Schwab2021}.  Mass loss during the giant phase would likely produce a H- and He-free nebula analogous to a standard planetary nebula (PN) around young WDs.   The central object might be dust-obscured during some of this evolution, particularly because of the carbon-enriched composition of the mass lost during the giant phase. \citet{Gvaramadze2019} reported the first detection of a possible WD merger remnant (IRAS 00500+6713) matching the above criteria:  a H-free nebula detected in WISE surrounding a H- and He-free star with a surface temperature around $2\times10^5$K and luminosity around $10^{4.5} L_{\odot}$. \citet{Gvaramadze2019} showed that IRAS 00500+6713 also had a remarkably high-speed wind suggestive of a strongly magnetized and rapidly rotating central object. Simulations of WD mergers indeed find that they can produce very strongly magnetized remnants \citep{Ji2013}. \citet{Ritter2021} later found high speed $\sim 1100$ km/s expansion of the Pa 30 nebula from the [S II] doublet. Strong mid-IR  \citep{Gvaramadze2019} and X-ray emissions \citep{Oskinova2020}, moderate [O III] 5007 \r{A} emissions, weak [Ar III] 7136 \r{A} features, and filamentary [S II] 6716 and 6731 \r{A} features \citep{Ritter2021, Fesen2023} from this nebula are also observed.

The first technique that we consider is whether merger candidates can be photometrically identified given their predicted luminosity and effective temperatures. This turns out to be challenging.  
The second technique we consider is more promising:  merger candidates can be spectroscopically identified by their unusual H-free photoionized nebulae. This is analogous to how the dusty circumstellar nebulae around R Coronae Borealis stars have proven valuable at elucidating the nature of these enigmatic stars \citep{DeMarco2002,Clayton2011,Clayton2011b,Montiel2018}.   In the case of more massive C/O merger remnants, however, we propose to use the surrounding nebula to infer the presence of a central merger remnant and we focus on the much hotter phase of the central remnant's evolution when the surrounding nebula is analogous to a photoionized planetary nebula.

To orient the reader, we note that the number of active C/O-C/O merger remnants in a galaxy can be estimated as follows: 
\begin{equation}
\label{eq:Nactive}
    {\rm N_{active}} \approx 10 \times \left(\frac{\dot{N}}{10^{-14}{\rm yr^{-1}}{\rm M_{\odot}}^{-1}} \frac{M_{\star}}{10^{11} {\rm M_{\odot}}}\frac{t_{\rm remnant}}{\rm 10^4 years}\right),
\end{equation}
where $\dot{N}$ is the WD merger rate, and $t_{\rm remnant}$ is the lifetime of a merger remnant. Assuming the super-Chandrasekhar double WD merger rate to be $ 1.0^{+1.6}_{-0.6} \times 10^{-14} {\rm yr^{-1}} {\rm M_{\odot}}^{-1}$ \citep{Badenes2012}, the number of white dwarf merger remnants in  the `thermal' phase of evolution highlighted in Figure \ref{fig:contaminants} is around 20 in the Milky Way or M31, and around 100 in M87. The total WD merger rate is probably $\sim 5-10 \times$ higher than the super-Chandrasekhar rate, though it remains uncertain what fraction of those mergers will explode as type Ia supernovae.

One general challenge in identifying WD merger remnants highlighted in Figure \ref{fig:contaminants} is that the remnants are similar in luminosity and effective temperature to intermediate-mass stars evolving to become WDs. In the Milky Way, sufficiently detailed observations may be possible to distinguish between these possibilities (e.g., by environment or spectroscopy).   In other galaxies, the natural way to minimize the impact of the contaminants in Figure \ref{fig:contaminants} is to focus on galaxies with high stellar masses (high WD merger rates) and low star formation rates (low birth rates of intermediate-mass stars), i.e., low specific star formation rates.  To be more quantitative, the birth rate of intermediate-mass stars that are the most likely to overlap in $L,T_{\rm eff}$ space with WD merger remnants is $\sim 10^{-2} \, {\rm yr^{-1}} \, (\dot M_\star/1 {\rm{\rm M_{\odot}} \, yr^{-1}})$.  Including sub-Chandrasekhar mass systems the WD merger rate is $\sim 5 \times 10^{-3} \, {\rm yr^{-1}} \, (M_\star/10^{11} \, \rm{\rm M_{\odot}})$.  Thus galaxies with specific star formation rates $\dot M_\star/M_\star \ll 5 \times 10^{-12}$ yr$^{-1}$ are likely to have low contaminants from intermediate-mass stars.  The focus of our work will thus be on massive, quiescent galaxies such as M87 or the bulge of M31.   

The remainder of this paper is organized as follows.  In \S \ref{sec:methods} we present our methods for constructing a composite stellar population for an old stellar population with both single and binary stellar isochrones as well as WD merger remnants.  We also summarize our methods for calculating photo-ionized nebulae surrounding WD merger remnants for comparison to planetary nebulae.   In \S \ref{sec:photometry}, we summarize the main observational challenges in identifying merger remnants based on photometric properties such as the luminosity and effective temperature of the central star. In \S \ref{sec:spectroscopy}, we show that spectroscopic signatures in the surrounding photoionized nebula are a more promising way of identifying WD merger remnants. Finally, in \S \ref{sec:discussions} we summarize our results and discuss optimal  strategies for identifying the unique population of WD merger remnants.  

\section{Methods}\label{sec:methods}
In \S \ref{sec:compCMD} we describe how we construct synthetic color-magnitude diagrams (CMDs) to assess the photometric detectability of WD merger remnants.   In \S \ref{sec:spectramethod} we discuss how we calculate the spectrum of the photoionized nebula surrounding WD merger remnants.
    
\subsection{Simulated Composite Color-Magnitude Diagram} \label{sec:compCMD}

In order to compare the theoretically predicted population of WD merger remnants with the rest of a galaxy's stellar population, we construct synthetic CMDs. We do so separately for single star and binary star models using the MESA Isochrones and Stellar Tracks code\footnote{\url{https://waps.cfa.harvard.edu/MIST/}}\citep[MIST,][]{MIST0,MIST1} and the Binary Population and Spectral Synthesis code\footnote{\url{https://bpass.auckland.ac.nz}}\citep[BPASS,][]{BPASS1,BPASS2,BPASS2.1}, respectively.

MIST self-consistently computes stellar evolutionary tracks using Module for Experiments in Stellar Astrophysics \citep[MESA,][]{Paxton2011, Paxton2013, Paxton2015, Paxton2018, Paxton2019}, covering a large parameter space in age, mass, and metallicity.  We use the python-Flexible Stellar Population Synthesis code\footnote{\url{https://dfm.io/python-fsps/}}\citep[python-fsps,][]{fsps1,fsps3} with MIST to construct a composite CMD. We assume a Salpeter initial mass function, because strong lensing suggests there is a significant population of low-mass stars in elliptical galaxies \citep{Leier2016}.   

The MESA models in Figure \ref{fig:contaminants} show that the primary single star contaminants come from the mass range between $6 <$ M $\rm < 10 \, {\rm M_{\odot}}$. This indicates that only single stars born in the past $\lesssim 300$ Myr will be in a similar part of the HR diagram  to our merger remnants. As a result, we take isochrones with $3 <$ log(age) $< 9 $ years with age bins of 0.5 in log. This increases the efficiency of generating our composite CMDs by eliminating the need to compute and plot numerous low-mass stars that are always fainter than our models of interest. Then, at each age, we interpolate the luminosity, effective temperature, fsps weights, and AB magnitudes at relevant wavebands to a linear grid of $10^6$ points plus an additional linear grid of $10^7$ points at luminosity $L>10^4 \rm L_{\odot}$ to cover the evolutionary track with enough resolution, including where stellar evolution is relatively rapid. To obtain the composite stellar population, we randomly sample from the interpolated luminosity and effective temperature with the probability distribution given by fsps. 
Since the contaminants here are from relatively massive stars, only the recent star formation rate affects the size of this population. Assuming a constant star formation rate in the past is thus a fine approximation.   Hence, at each age bin, we sample a total of $N = \Delta m/m_{\rm mean} \simeq (\dot M_{\star}/0.1 {\rm M_{\odot}}) \times \Delta t$ stars, where $m_{\rm mean}$ is the mean stellar mass of the IMF, which we set to 0.1${\rm M_{\odot}}$ here. 

It is well known that mass transfer and stellar mergers in binary systems can make an older stellar population appear bluer than predicted by single stars alone.   This is believed to be the origin of the UV upturn in elliptical galaxies \citep{Yi2004, Han2007}.  Binary evolution is thus likely to introduce additional sources that appear similar to WD merger remnants on top of those associated with single intermediate-mass stars.    Since the MIST isochrones do not evolve binary systems, we alternatively adopt  BPASS  models \citep{BPASS1,BPASS2,BPASS2.1} to quantify the impact of binarity on our model CMDs. With binary evolution, stars with the same mass and age can be inherently different, so using python-fsps is no longer an option for sampling a CMD as it only has access to the integrated SSP spectra from BPASS. Therefore, we use the Hoki code to access and sample BPASS models\footnote{\url{https://heloises.github.io/hoki/}} \citep{Hoki}.

With Hoki, we start with a BPASS model and sample stellar populations in a range of log ages $6 <$ log(age [yr]) $< 9 $, leaving out low-mass stars like before. For each value of luminosity and effective temperature in a $i\times j$ grid, Hoki provides a weight $W_{i,j}$ that indicates the number of systems formed in a single 1 million ${\rm M_{\odot}}$ population. As a result, the probability of a system being present in a population is given by $\mathcal{P}_{i,j} = W_{i,j}/\sum W_{i,j}$. As with MIST, we sample a total of $N \simeq (\dot M_{\star}/0.1 {\rm M_{\odot}}) \times \Delta t$ number of stars based on this probability to construct the composite stellar population. Each sample includes a system's luminosity and effective temperature on the grid, but lacks information about the initial stellar mass of the system.   We also note that BPASS reports the properties of  the primary and secondary stars separately and independently; the connection between the two stars in a binary system has not yet been implemented.

    \begin{figure*}
        \centering
        \includegraphics[width=\columnwidth]{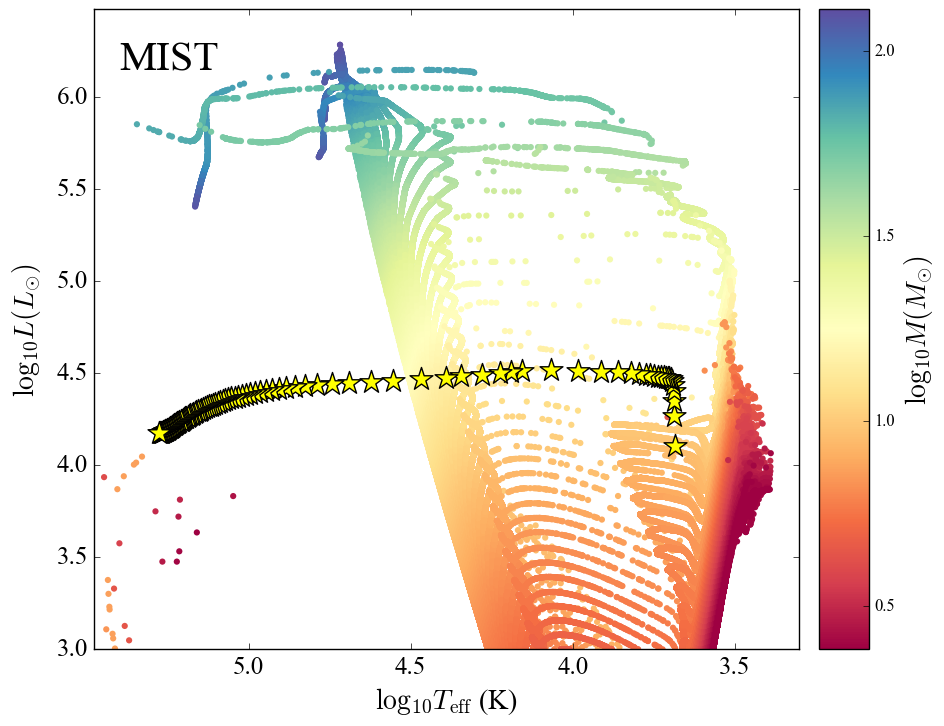}
        \includegraphics[width=0.93\columnwidth]{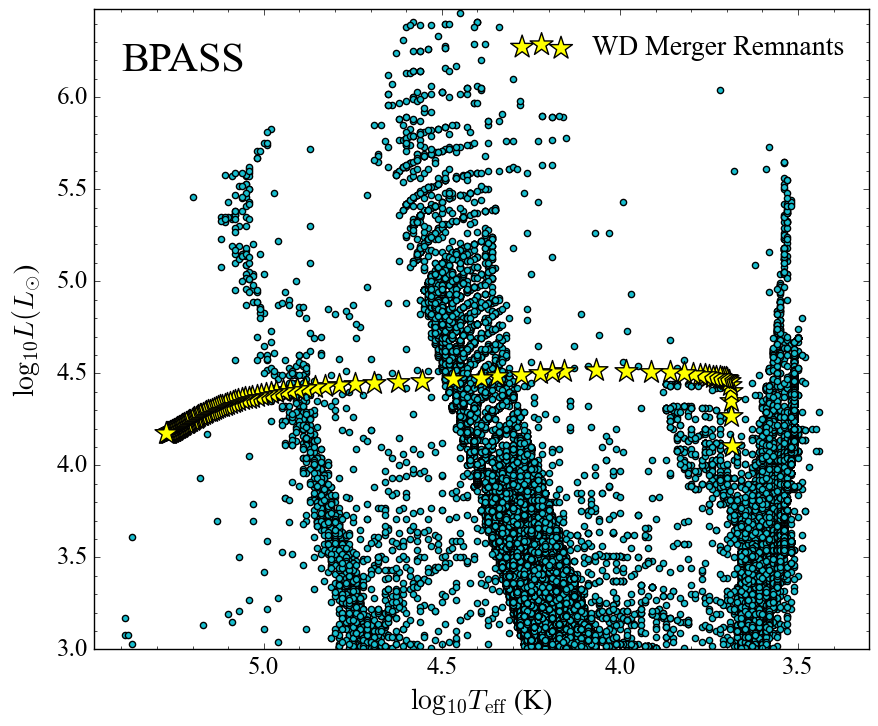}
        \caption{Simulated H-R diagram of a massive galaxy with a low star formation rate and constant star formation history sampled from MIST (left) and BPASS (right) isochrones. The luminosity of the WD merger remnants from \citet{Schwab2016} are plotted in star-shaped points; the population is estimated to be around 70 for a galaxy with the stellar mass of M87 ($\rm 6.8 \times 10^{11} {\rm M_{\odot}}$ in \citealt{Forte2012}). The color in the left panel indicates the stars' initial masses, but we do not have that information for BPASS models, as systems with different masses can appear in the same part of the HR diagram due to binary interaction.}
        \label{fig:HR}
    \end{figure*}

\subsection{Simulated Nebula Spectrum}
\label{sec:spectramethod}

Mass loss may play an important role in the thermal evolution of the merger remnant, particularly during the giant phase when the surface gravity of the star is low, reaching an escape velocity of $\rm \lesssim 60 km/s$.   Such ejected mass could extend to $\rm \sim 2\times 10^{18} cm$ as the remnant evolves over its $\sim 10^4$ years of thermal evolution.  The winds are likely to be dusty due to the remnants' C/O-dominated composition. They may appear bright in the infrared because of dust reprocessing \citep{Schwab2021}.  Here we focus on using the CLOUDY spectral synthesis code\footnote{\url{https://gitlab.nublado.org/cloudy/cloudy}} \citep[v17.02; ][]{CLOUDY1998,Ferland2017} to calculate the spectrum of the surrounding nebula accounting for photoionization and dust. 

We use CLOUDY to perform a suite of photoionization simulations both for our WD merger remnant with a C/O-dominated composition, and for normal planetary nebulae with a solar composition.  We  assume a total ejecta mass $M_{\rm ejecta} = 0.1 {\rm M_{\odot}}$ based on \citet{Schwab2021}, although we note that the exact ejecta mass is uncertain (calculations with ejecta masses of $0.05 {\rm M_{\odot}}$ and $0.20 {\rm M_{\odot}}$ were very similar to the models shown here).    
We assume blackbody emission from the central source and consider a range of ejecta sizes and stellar effective temperatures. 
Since the luminosity of the merger remnant remains relatively constant between $\rm 10^4 L_{\odot}$ and $\rm 3 \times 10^4 L_{\odot}$ (see Figure \ref{fig:contaminants}), we use the same luminosity $\rm \sim 5\times10^{37} erg s^{-1}$ in all our simulations. Tests using $\sim 2 \times$ higher luminosity produce similar results. 

To obtain the abundance of elements for the WD merger remnant, we use a MESA test suite which produces a $\rm 0.6 {\rm M_{\odot}}$ carbon-oxygen WD, roughly the expected mass of one WD in the binary system, from a $\rm 3 {\rm M_{\odot}}$ progenitor. We take the final mass fraction of elements in this white dwarf as the composition of the ejecta, which includes C, N, O, and Ne, and trace amounts of H and He. Additionally, we assume solar abundance for elements heavier than neon, including Mg, Si, S, Cl, Ar, and Fe. Based on the calculations of \citet{Schwab2016} (their Figure 10), we expect that this assumption is valid throughout the majority of the thermal evolution of the merger remnant; only at the latest stages, when a convectively bounded Ne flame develops, does the surface abundance change significantly (at least according to these spherically symmetric models).  

Since the composition of the ejecta is C/O-rich, a large amount of dust is expected to be produced during the giant phase when the photosphere is cool. As a result, modeling the effect of dust in CLOUDY is crucial to understanding the appearance of the merger remnant. Our default models with dust assume that 10\% of the ejecta by mass is graphite grains. CLOUDY performs the heating and cooling of dust grains self-consistently, and models the size distribution using 10 bins. Such dust can survive outside the sublimation radius 
\begin{equation}
    R_{\rm sub} = \left(\frac{L}{4\pi \sigma T_{\rm sub}^4}\right)^{1/2} 
\end{equation}
where $R_{\rm sub} \simeq 10^{14} \rm cm$ when $T_{\rm sub} \rm \sim 1500 K$. Since $R_{\rm sub}$ is orders of magnitudes smaller than the outer radius of the ejecta, dust grains will very likely be present in the ejecta.

\section{Photometric Detectability of Merger Remnants}\label{sec:photometry}

We begin by discussing the detectability of merger remnants in theoretical $L-T_{\rm eff}$ space and then consider the more observationally relevant case of optical and UV CMDs.    We adopt properties of an M87-like elliptical galaxy with $\rm SFR = 0.01 {\rm M_{\odot}}/yr$ \citep{Davis2014}. In between $3 <$ log(age) $< 9$ years, we sample a total of $10^8$ stars that follows a Salpeter-like initial mass function. For BPASS, we sample the same number of stars in each corresponding age bin. We use the stellar mass of M87  of $\rm 6.8 \pm 1.1 \times 10^{11} {\rm M_{\odot}}$ \citep{M87Mass} to estimate a total of $70$ active super-Chandrasekhar remnants; we thus sample a total of 70 systems from \citet{Schwab2016}'s MESA model with even spacing for the remnant's age during its evolutionary track.   We note that our results are converged with respect to taking a wider age range for sampling the isochrones because the older stars included by doing so only appear in the lower luminosity parts of the HR diagram.   We also reiterate that sub-Chandrasekhar merger remnants are initially similar in luminosity and effective temperature to super-Chandrasekhar merger remnants (Figure \ref{fig:contaminants}), which likely increases the number of expected remnants relative to the 70 sampled here and shown in Figures \ref{fig:HR} and \ref{fig:CMD}.   

\subsection{Theoretical Distinction in \texorpdfstring{$L-T_{\rm eff}$}{} Space}

Figure \ref{fig:HR} shows mock H-R diagrams for our M87-like galaxy from MIST and BPASS.  In both cases, the general distribution of stars is very similar in the $T_{\rm eff} \lesssim 10^{4.7}$ K part of the H-R diagrams, which represent the main sequence and giant phases.  On the MIST plot, we see a small number of stars in the proto-WD/planetary nebula phase. These are the main sources of contamination in $L-T_{\rm eff}$ space. However, the white dwarf merger remnants are typically a few times more luminous and significantly more abundant for a galaxy like M87.  However, the number of standard single stars in this portion of the HR diagram will scale up linearly with the star formation rate. Hence, in a Milky Way-like star-forming galaxy with $\rm SFR \sim 1 {\rm M_{\odot}}/yr$, these contaminants will significantly outnumber the white dwarf merger remnants (which will also be less numerous because of the lower total stellar mass of the Milky Way). In this case, additional information is needed to distinguish between the photometrically similar objects, such as the spectroscopic features described in the next section. For the bulge of the Milky Way or M31, on the other hand, the results in Figure \ref{fig:HR} will remain qualitatively applicable, though the total number of WD mergers will be smaller because of the lower total stellar mass of the system (eq. \ref{eq:Nactive}).

In the BPASS models in the right panel of Figure \ref{fig:HR} there is a significant population of stars on the hotter part of the H-R diagram, which is absent in the single-star evolutionary tracks. These are   stars whose hydrogen envelopes have been stripped during binary evolution. The lower luminosity stars are helium stars while the higher luminosity stars are primarily Wolf-Rayet stars.   
The existence of this hotter population of stripped stars somewhat 
complicates the identification of white dwarf merger remnants in  L-$\rm T_{eff}$ space, but the merger remnants still stand out at $T_{\rm eff}\gtrsim \rm 10^5 K$. 

\subsection{Observational Distinguishability}

    \begin{figure*}
        \centering
        \includegraphics[width=\columnwidth]{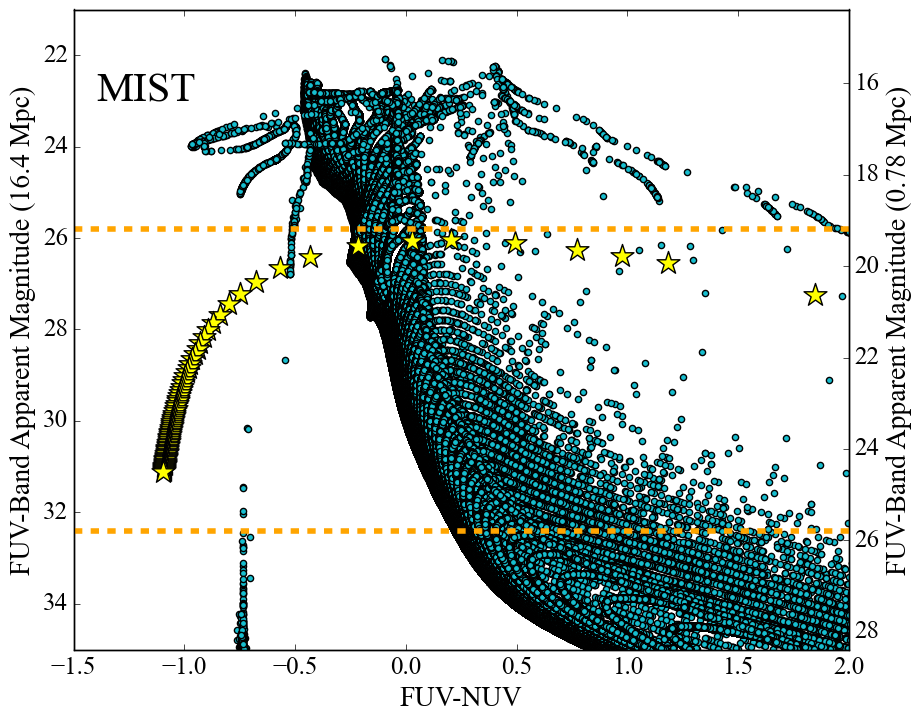}
        \includegraphics[width=\columnwidth]{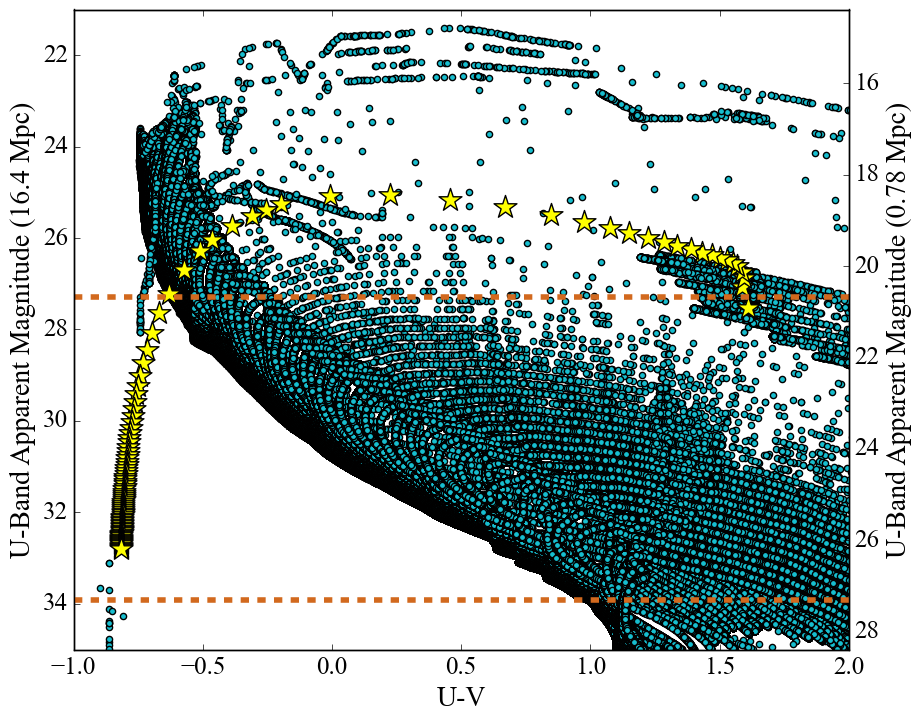}\\
        \includegraphics[width=\columnwidth]{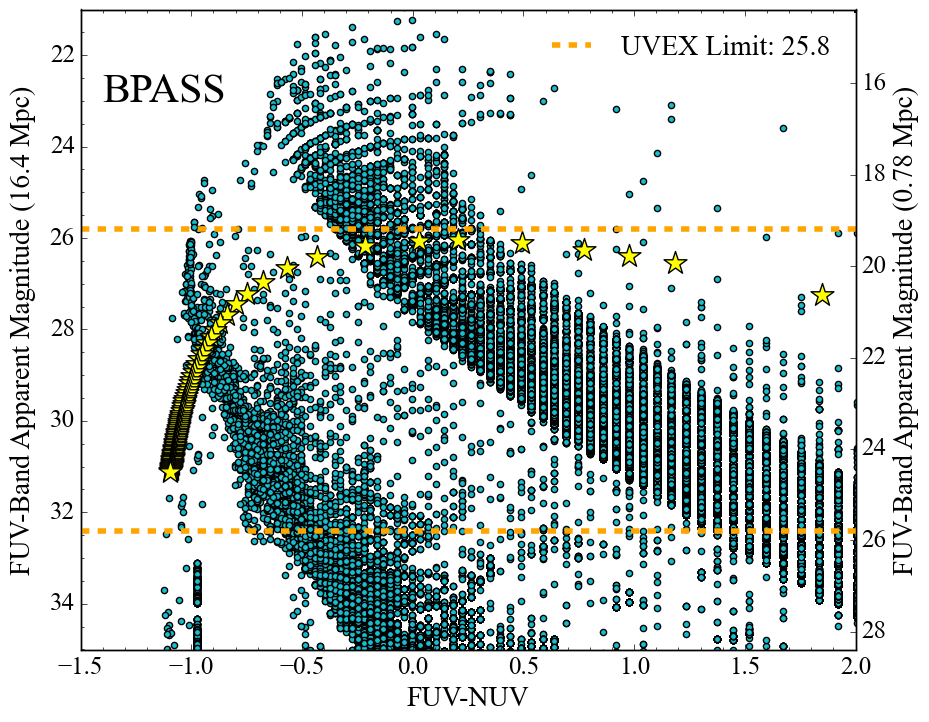}
        \includegraphics[width=\columnwidth]{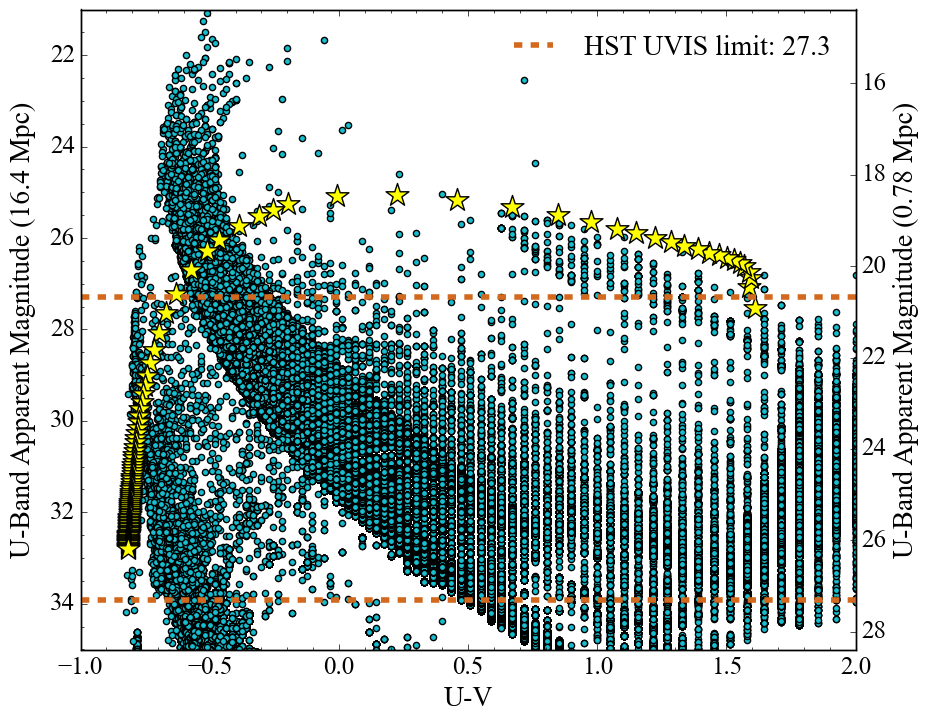}        
        \caption{CMDs derived from the H-R diagrams in Figure \ref{fig:HR} for MIST (top) and BPASS (bottom) isochrones. Stars (blue dots) are sampled from the isochrones and WD merger remnants are from \citet{Schwab2016}. The simulated plots are in FUV (1565 \r{A}) and NUV (2740 \r{A}) filters (left), or the U (3640 \r{A}) and V (5400 \r{A}) filters (right). 
        The dotted line highlights the observational limiting sensitivity of The Ultraviolet Explorer (UVEX, orange) and the Hubble Space Telescope's ultraviolet and visible light imager (HST UVIS, brown) when placing the galaxy at two different distances (16.7 \& 0.78 Mpc).
        The bottom panel highlights that existing and planned UV filters are not blue enough to distinguish He stars (the blue ridge of BPASS points) and WD merger remnants.}
        \label{fig:CMD}
    \end{figure*}

We now consider how single \& binary stars and WD merger remnants would appear in observationally accessible ultraviolet (FUV \& NUV) and optical (U \& V) bands. Python-fsps computes the AB magnitude of stars  for a large number of commonly used filters. We choose the FUV and NUV filters centered on 1565 \r{A} and 2740 \r{A}, and the U and V filters centered on 3640 \r{A} and 5400 \r{A} respectively. For the white dwarf merger remnant or stellar systems sampled from BPASS, we assume blackbody emission to compute their AB magnitudes. Note that for the WD merger remnants, this assumption neglects the impact of the surrounding ejecta, which will be considered in the following section. 

The CMDs from MIST and BPASS in both the optical and UV filters are plotted in Figure \ref{fig:CMD}. For the MIST CMDs, we notice that the merger remnants easily stand out from other stars at later stages of evolution, particularly in the UV bands.  However,  the limiting sensitivity of the most advanced near-term UV instrument, the Ultraviolet Explorer (UVEX), is 25.8 mags \citep{Kulkarni2021}.   In this case, Figure \ref{fig:CMD} shows that merger remnants cannot be detected in the UV at the distance of M87 (16.4 Mpc). A limiting sensitivity closer to 28 mags or, equivalently, observing a closer galaxy at a distance of 10.9 Mpc such as NGC 1023, is required for WD merger remnants to be detectable in the UV in the left part of the CMD. Likewise, the right top panel of Figure \ref{fig:CMD} suggests that deep HST imaging of a galaxy significantly closer than M87 (such as Andromeda at 0.78 Mpc indicated by the lower dotted lines) might be sufficient to detect some of the merger remnants at high effective temperature.

The bottom panels of Figure \ref{fig:CMD} show the analogous UV and optical CMDs for BPASS models.   The He stars are now a major contaminant at essentially all accessible UV or optical luminosity.  The reason is simply that an FUV filter of 1565 \r{A} is still not short enough wavelength to be sensitive to the higher effective temperature of the WD merger remnants relative to He stars.  Overall, given realistic binary stellar populations, Figure \ref{fig:CMD} strongly suggests that deep photometric imaging is unlikely to be an effective route to identifying candidate WD merger remnants even in the case of nearby galaxies such as the bulge of M31.

A caveat in the analysis presented here is that Hoki provides only the characteristics of primary or secondary stars separately, but there is no information about which secondary and primary stars go together.   Hence, all the models plotted here are the primary stars of each binary system, which corresponds to the hotter star. This is true for the CMDs in Figure \ref{fig:CMD} as well as the BPASS models in the H-R diagram (Figure \ref{fig:HR}). Since the He stars are likely much hotter than any main sequence companion, the UV CMD should be accurate. However, it is possible that in the optical the secondary will dominate the system's luminosity in which case the BPASS optical CMD would likely be more similar to the MIST CMDs.   Even in this case, the WD merger remnants are still very difficult to detect. Additionally, we will discuss how dust will affect the transmitted stellar continuum, and hence the UV and optical luminosity, in \S \ref{sec:dust}.

    \begin{figure*}
        \centering
        \includegraphics[width=\textwidth]{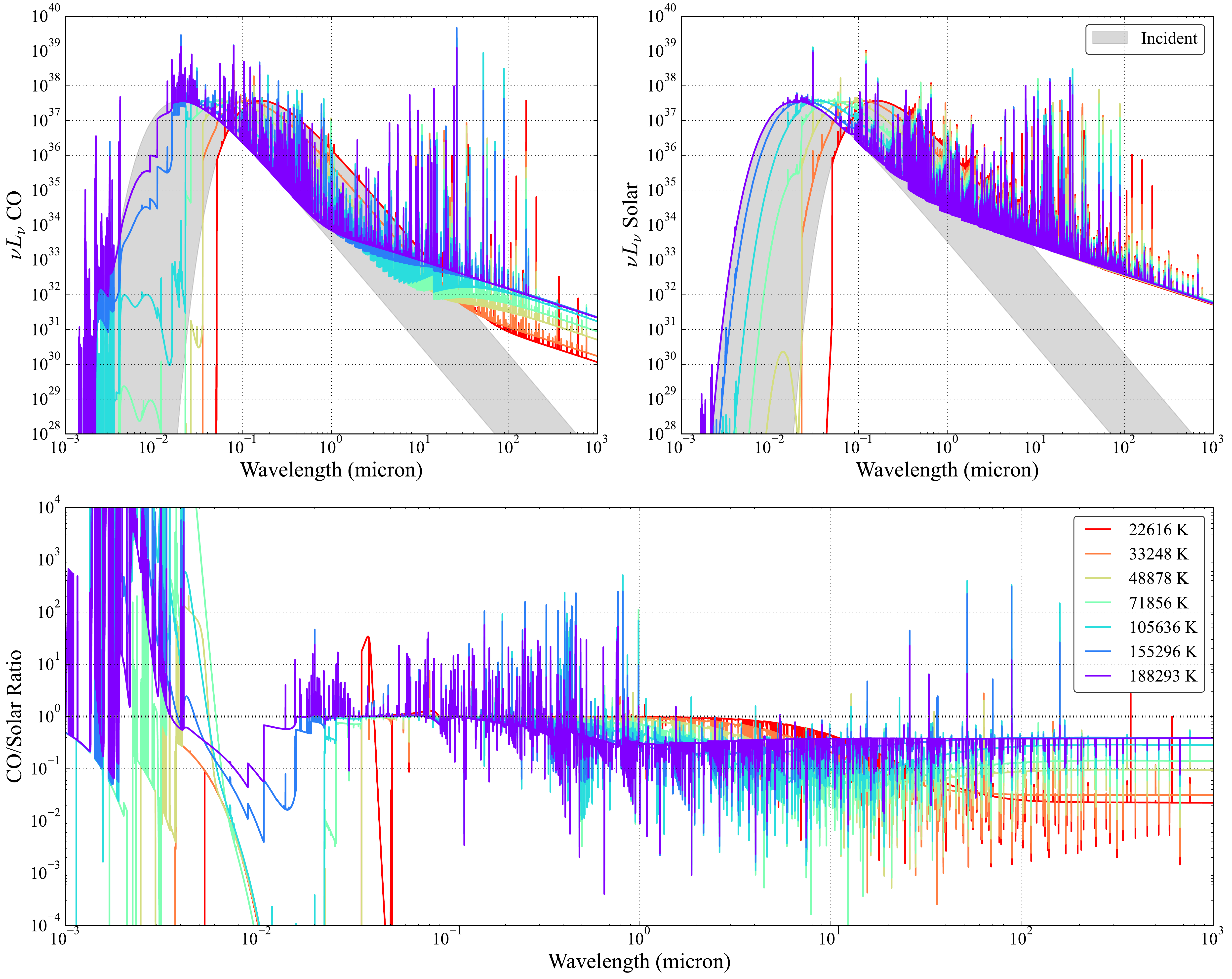}\\
        \caption{Output spectra from CLOUDY for C/O (left) and solar (right) composition nebula, with the former representing our predictions for WD merger remnants.  The bottom panel highlights the ratio between the two spectra. We fix the inner radii of the clouds here to be $\rm 10^{17} cm$, while varying the effective temperature of the central star between $\rm \sim 2\times10^4 - 2\times10^5 K$. The grey shaded region represents the blackbody spectrum from the range of temperature in the legend. No dust is included in these models; see Figure \ref{fig:dust} for models with dust.}
        \label{fig:full_spec}
    \end{figure*}

\subsection{Novae as Contaminants}

Classical novae can also appear in the same region of the H-R diagram as WD merger remnants and are likely an even more significant contaminant than He stars.
Classical novae are binary systems hosting a WD accreting from a non-degenerate stellar companion.  The accretion leads to a thermonuclear runaway on the surface of the WD which causes the WD to brighten to near Eddington and then cool and contract over time (see \citealt{Chomiuk2021} for a review).  The process repeats with a recurrence time that is a function of the mass of the WD and the accretion rate.

Figure \ref{fig:contaminants} shows this  post-outburst nova cooling phase for $0.6$ and $\rm 1{\rm M_{\odot}}$ WDs (using models from \citealt{Wolf2013}).   The time between points is 1 yr and 30 yr for the $\rm 1{\rm M_{\odot}}$ and $\rm 0.6{\rm M_{\odot}}$ models, respectively.   Figure \ref{fig:contaminants} shows that post-outburst novae appear very similar to WD merger remnants.   Moreover, the cooling timescale for the post-outburst novae  is sufficiently long that secular cooling is unlikely to distinguish them from WD merger remnants in all cases.   The most promising part of parameter space for distinguishing post-outburst novae and WD merger remnants photometrically is the hottest most luminous systems because the post-outburst novae evolve on $\sim$ year timescales.

We can estimate the number of potential novae contaminants using the estimated novae rate of $26\pm 5$ in the Milky Way \citep{Kawash2022} and $363^{+33}_{-45}$ in M87 \citep{Shafter2017b}. The year- to decade-long evolution timescale in Figure \ref{fig:contaminants} for novae translates to $\sim 10-100$ years as contaminants. Hence, there are likely $\sim 10^4 - 10^5$ post-outburst novae in M87 and  $\sim 10^3 - 10^4$ in the Milky Way, $\sim 10-100$ times more abundant than WD merger remnants.

Despite the photometric resemblance between post-outburst novae and WD merger remnants, the spectroscopic differences between them should be significant.  In particular, novae eject a $\sim 10^{-4-5} {\rm M_{\odot}}$ envelope per eruption (with significant hydrogen), while WD merger remnants are likely  surrounded by $\sim 0.1 {\rm M_{\odot}}$ of ejecta with a very different composition.   In the next section, we quantify the spectroscopic signatures of the photoionized nebulae surrounding WD merger remnants. 

\section{Spectroscopic Signatures of WD Merger Remnants}\label{sec:spectroscopy}

\subsection{CLOUDY spectra}\label{sec:nodust}

    We ran a 2D logarithmically spaced grid (20x10) of simulations across a range of stellar effective temperatures and ejecta radii expected for the merger remnants: from $\sim 5 \times 10^3$K to $\sim 2 \times 10^5$K, and  inner radii range between $10^{16}$cm and $10^{19}$cm.  We consider both C/O and solar compositions in order to compare merger remnant nebulae and standard planetary nebulae.  To begin the models do not include dust but we return to models with dust in \S \ref{sec:dust}.   It is worth noting that the first ionization energy of oxygen is nearly identical to that of hydrogen so that the ionization of oxygen and hydrogen set in at very similar stellar effective temperatures.
    
    CLOUDY outputs the incident and transmitted spectrum as in Figure \ref{fig:full_spec}, allowing us to identify spectral signatures of nebulae produced by WD merger remnants.  The shaded region in Figure \ref{fig:full_spec} reflects the blackbody incident continuum for the range of effective temperatures in the legend.  The output continuum emission at long wavelengths in Figure \ref{fig:full_spec} is free-free emission from the photo-ionized nebula.
    In the soft X-ray ($\sim 10^{-3}-10^{-2}$ microns), the C/O remnants produce more prominent emission features. In the extreme UV ($\sim 10^{-2}-10^{-1}$ microns), the cloud is more opaque to incident radiation for the C/O remnant until the central star becomes hotter than $\rm \sim 10^5 K$. 
    
    Simulated CLOUDY spectra also enable us to take the ratio of line strengths between individual spectral lines produced by the two photo-ionized clouds. For all parameters, almost all hydrogen lines in the C/O-rich remnant are more than three orders of magnitudes weaker than in the solar remnant. This is of course due to the lack of hydrogen in the merger remnant ejecta; hence the typical PNe $\rm H \alpha$ and $\rm H \beta$ line features should be extremely weak or nonexistent in WD merger remnants. 
    
\begin{figure*}
        \centering
        \includegraphics[width=\textwidth]{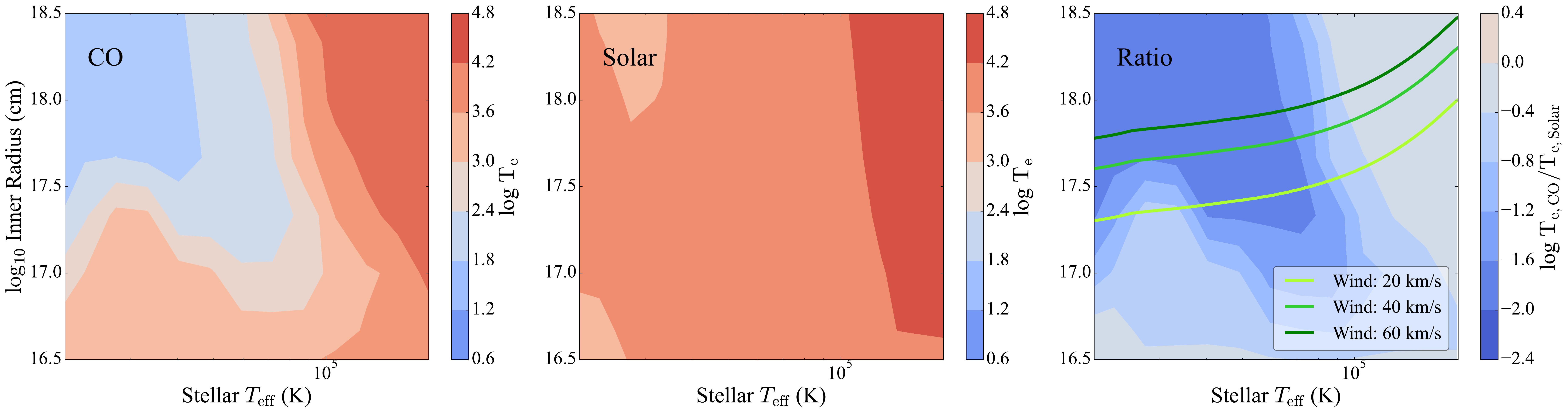}
        \caption{Electron temperature of the photoionized nebula from CLOUDY for the C/O (left) and solar (middle) composition clouds. The right panel highlights the temperature ratio showing that the effectively high metallicity C/O nebulae are significantly cooler than standard planetary nebulae;  overplotted is the expected trajectory in the evolution of WD merger remnants for outflow velocities of 20, 40, and 60 km/s.}
        \label{fig:temperature}
    \end{figure*}

\begin{figure*}
        \centering
        \includegraphics[width=0.327\textwidth]{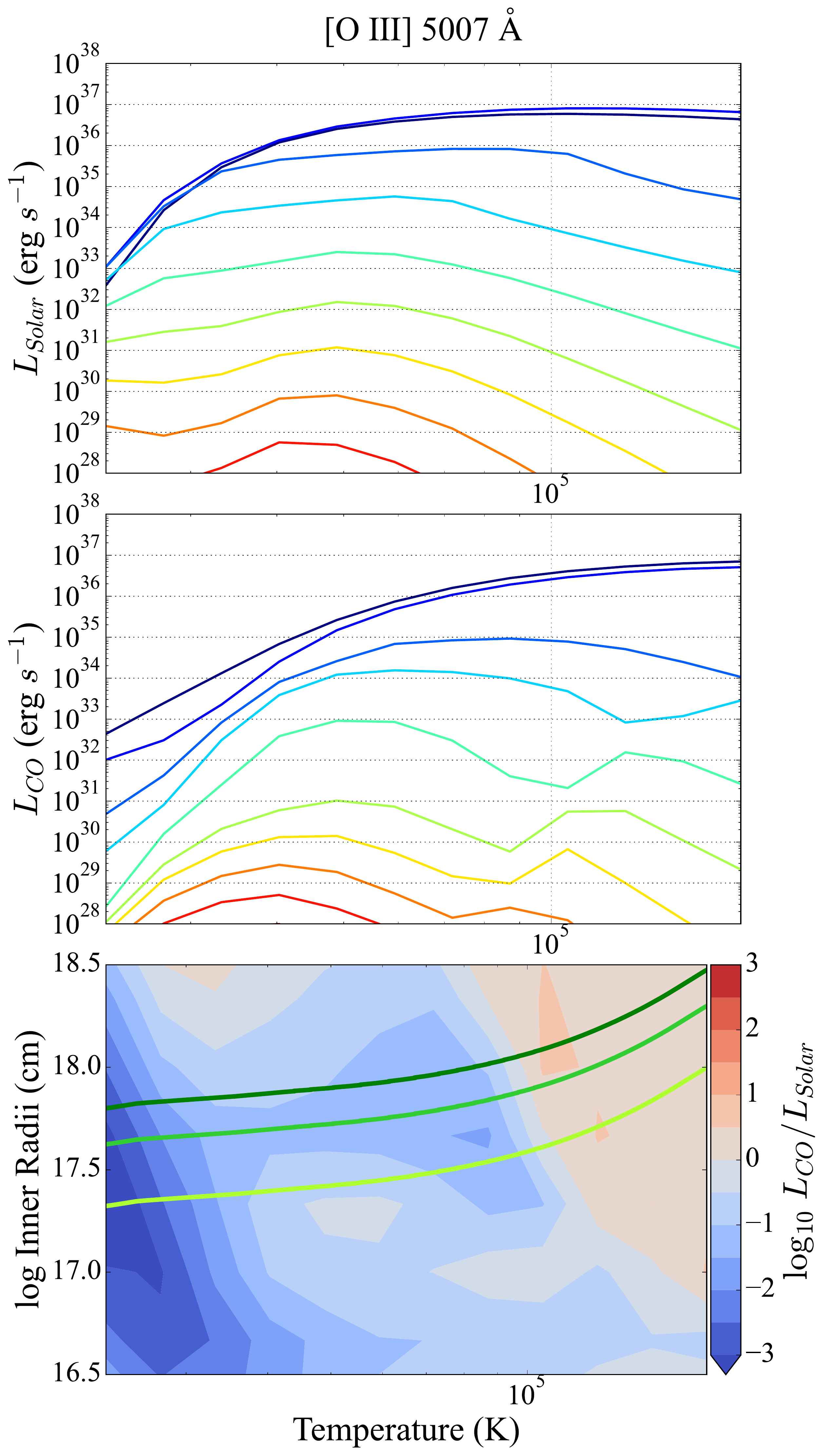}
        \includegraphics[width=0.31\textwidth]{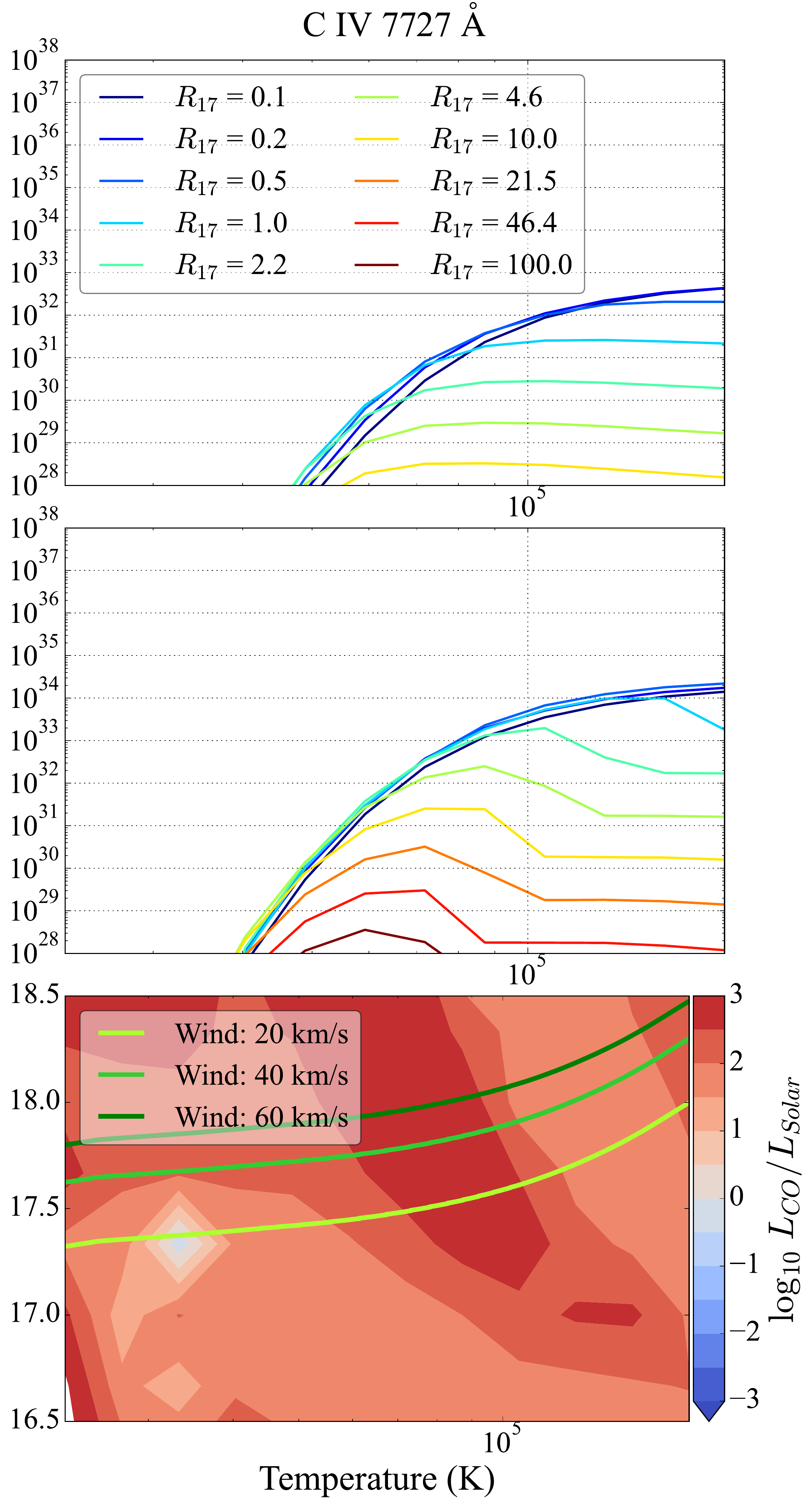}
        \includegraphics[width=0.31\textwidth]{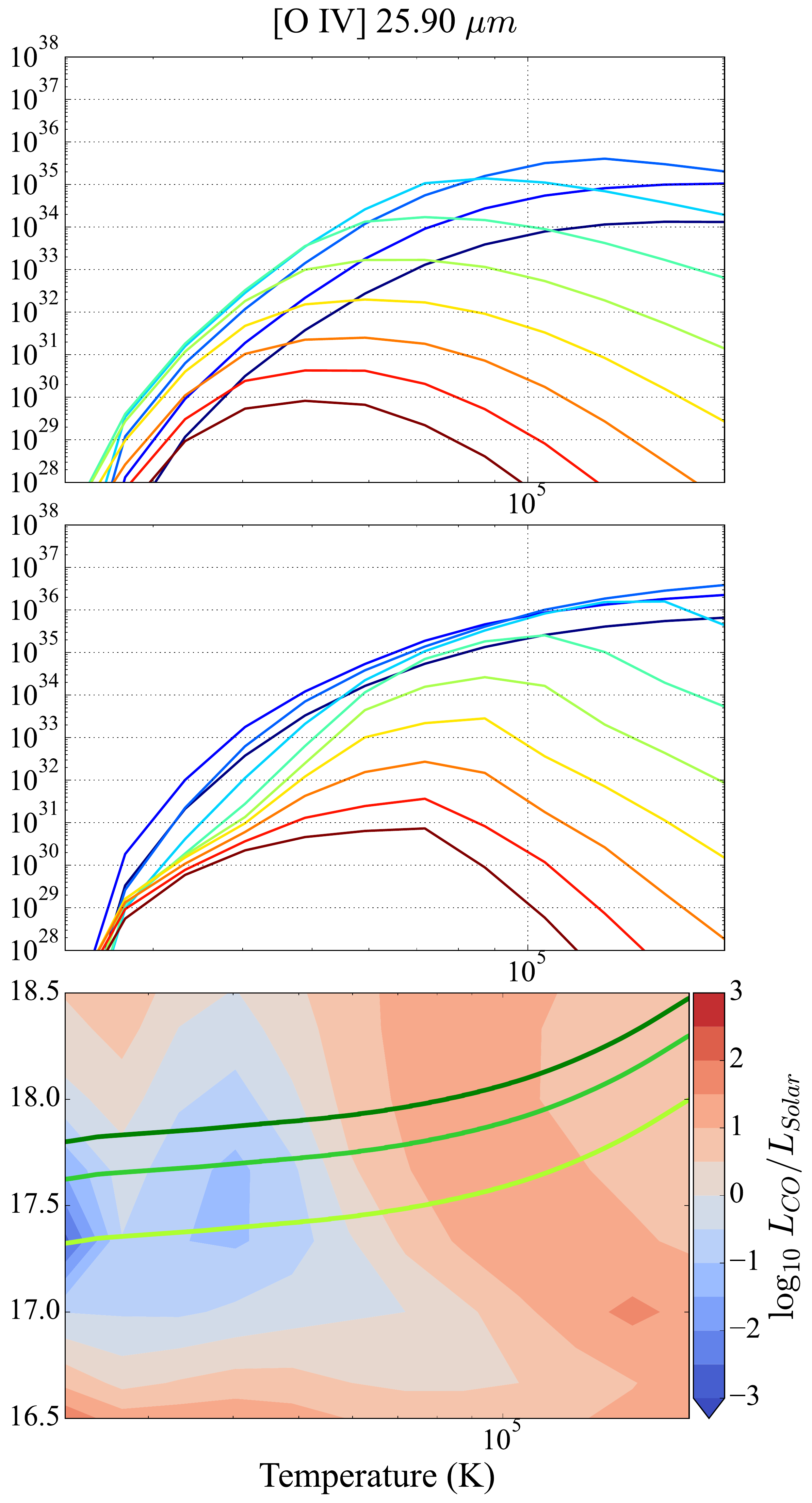}
        
        \caption{Select line luminosity for solar (top) and C/O (middle) composition nebulae in a grid of simulations across a range of stellar effective temperatures and cloud sizes ($R_{17}$ is cloud inner radii in units of $10^{17}$cm). The bottom panel highlights the ratio of the line luminosity overplotted with the expected trajectory in the evolution of WD merger remnants for outflow velocities of 20, 40, and 60 km/s. We predict that WD merger remnants are bright in [O III] 5007 \r{A}, C IV 7727 \r{A} and [O IV] 25.8832 $\rm \mu m$ but faint in hydrogen lines.}
        \label{fig:spec_grid}
    \end{figure*}

    Since our target objects are C/O-rich, we then focus on carbon and oxygen emission lines in the output spectra. Planetary nebulae are known for bright [O III] 5007 \r{A} lines where oxygen is doubly ionized above stellar effective temperatures of 30,000 K. Hence, this line is often used in planetary nebula surveys. The significantly larger abundance of oxygen in the merger remnant does not necessarily guarantee a much brighter [O III] 5007 \r{A} line. Unlike standard planetary nebulae in which oxygen is the dominant coolant, the WD merger remnants effectively have much higher metallicity, allowing the ejecta to cool more rapidly. Figure \ref{fig:temperature} shows that the C/O dominated nebula has significantly lower electron temperatures than solar composition nebula in almost the whole parameter space we have explored. Since collisionally excited lines are 
    weaker at lower temperatures, even though the WD merger remnant is rich in oxygen, the [O III] 5007 \r{A} line can be similar in strength, or even weaker than, in a typical solar composition PNe.   This is shown explicitly in Figure \ref{fig:spec_grid} which compares the strength of the [O III] 5007 \r{A} line, the C IV 7727 \r{A} line, and the [O IV] 25.9 $\rm \mu m$ line in solar composition and CO-dominated nebulae. We note that this carbon recombination line is rarely observed from photoionized gas, which could be a unique tracer for WD merger remnants. It is, however, also faint and so likely only detectable in the MW. In the bottom row, the lines show radii vs. stellar effective temperature given the post-merger models of $T_{\rm eff}$ vs. time in Figure \ref{fig:contaminants} and assuming outflow velocities of 20, 40, and 60 km/s.
    
    The left panels of Figure \ref{fig:spec_grid} show that in the later stages of the merger remnant, the [O III] line-strengths in solar and C/O composition ejecta are typically within a factor of $\sim 10$ of each other.  Note that these larger ejecta radii are   where the system  spends most of its time and where dust has the least impact.     Hence, an initial search for potential WD merger remnant candidates can identify planetary nebulae with strong [O III] 5007 \r{A} lines and weak hydrogen lines.  Table \ref{tab:lines} provides a list of additional spectral lines that are the strongest for the merger remnant, when compared to typical solar PNe.
    Spectral features that will potentially help narrow down the number of candidate WD merger remnant nebulae include bright carbon and oxygen recombination lines, as well as oxygen fine structure lines.  The middle and right panels of Figure \ref{fig:spec_grid} show, e.g., that the C IV 7727 \r{A} line is typically $\sim 40$  times stronger in the C/O-rich nebula and 
    the [O IV] 25.8832 $\rm \mu m$ line is $\sim 10$ times stronger when the effective temperature of the C/O-rich merger remnant is larger than $10^5$ K. Other oxygen fine-structure lines including [O III] 51.81 $\rm \mu m$ and [O III] 88.36 $\rm \mu m$ are also significantly stronger, but lie outside the range of most current instrumentation, notably JWST.  Many other oxygen and carbon lines in the UV or optical are also suitable for the same purpose, especially carbon and oxygen recombination lines such as C IV 4657 \r{A} and O IV 4631 \r{A}.   
    %The horizontal dotted lines in the left and middle panels of Figure \ref{fig:spec_grid} show the line sensitivity of the MUSE integral field spectrograph at the distance of M87.  Until the nebula expands to very large radii where it becomes highly ionized, many of the diagnostic lines should be detectable.    

\renewcommand{\arraystretch}{1.2}
\begin{table}
    \centering
        \begin{tabular}{|c|c|c|c|}
            \hline
            Line & Wavelength & $F_{\rm WD}/F_{\rm PN}$ & $F_{\rm WD,d}/F_{\rm PN,d}$\\
            \hline
            \hline
            \multicolumn{4}{|c|}{UV}\\
            \hline
              C IV & 2528.83 \r{A}  & 34.2  & 40.2\\
              O IV & 2450.01 \r{A}  & 22.2  & 26.0\\
              O IV & 3037.66 \r{A}  & 20.9  & 25.5\\
              O  V & 2941.99 \r{A}  & 31.3  & 29.6\\
            \hline
            \multicolumn{4}{|c|}{Optical}\\
            \hline
              C IV & 4658.74 \r{A}  & 34.9  & 41.0\\
              C IV & 7727.12 \r{A}  & 35.5  & 41.7\\
              O IV & 4632.41 \r{A}  & 22.9  & 26.8\\
              O IV & 7714.25 \r{A}  & 23.6  & 27.4\\
              O  V & 4930.35 \r{A}  & 32.6  & 30.4\\      
              O  V & 7612.08 \r{A}  & 34.1  & 32.4\\          
              O  V & 6486.60 \r{A}  & 25.6  & 25.7\\    
              O VI & 5291.36 \r{A}  & 78.1  & 76.7\\       
            \hline
            \multicolumn{4}{|c|}{IR}\\
            \hline
              [O III] & 88.3585 $\rm \mu m$  & 2.2  & 7.5\\
              $\left[\rm O\,III \right]$ & 51.8184 $\rm \mu m$  & 1.2  & 4.1\\
              $\left[\rm O\,IV \right]$ & 25.8832 $\rm \mu m$  & 5.7  & 12.2\\
              $\left[\rm O\,V \right]$ & 32.6035 $\rm \mu m$  & 5.0  & 6.4\\
            \hline
        \end{tabular}
        \caption{Notable emission lines from CLOUDY photoionization calculations for the C/O WD merger remnant. The flux ratios ($F_{\rm WD}/F_{\rm PN}$) are for comparison with a typical solar-composition PNe with the same stellar and ejecta attributes: a stellar effective temperature of 150,000 K, and an ejecta shell with a mass of $\rm 0.1{\rm M_{\odot}}$ and an inner radius of $10^{17}$ cm, where both models are dust-free. $F_{\rm WD,d}$ represents a model merger remnant nebula with a D/G mass ratio of 10\%, while $F_{\rm PN,d}$ has a D/G mass ratio of 1\%.}
        \label{tab:lines}
\end{table}

\subsection{Models With Dust}
\label{sec:dust}

    \begin{figure*}
        \centering
        \includegraphics[width=1\columnwidth]{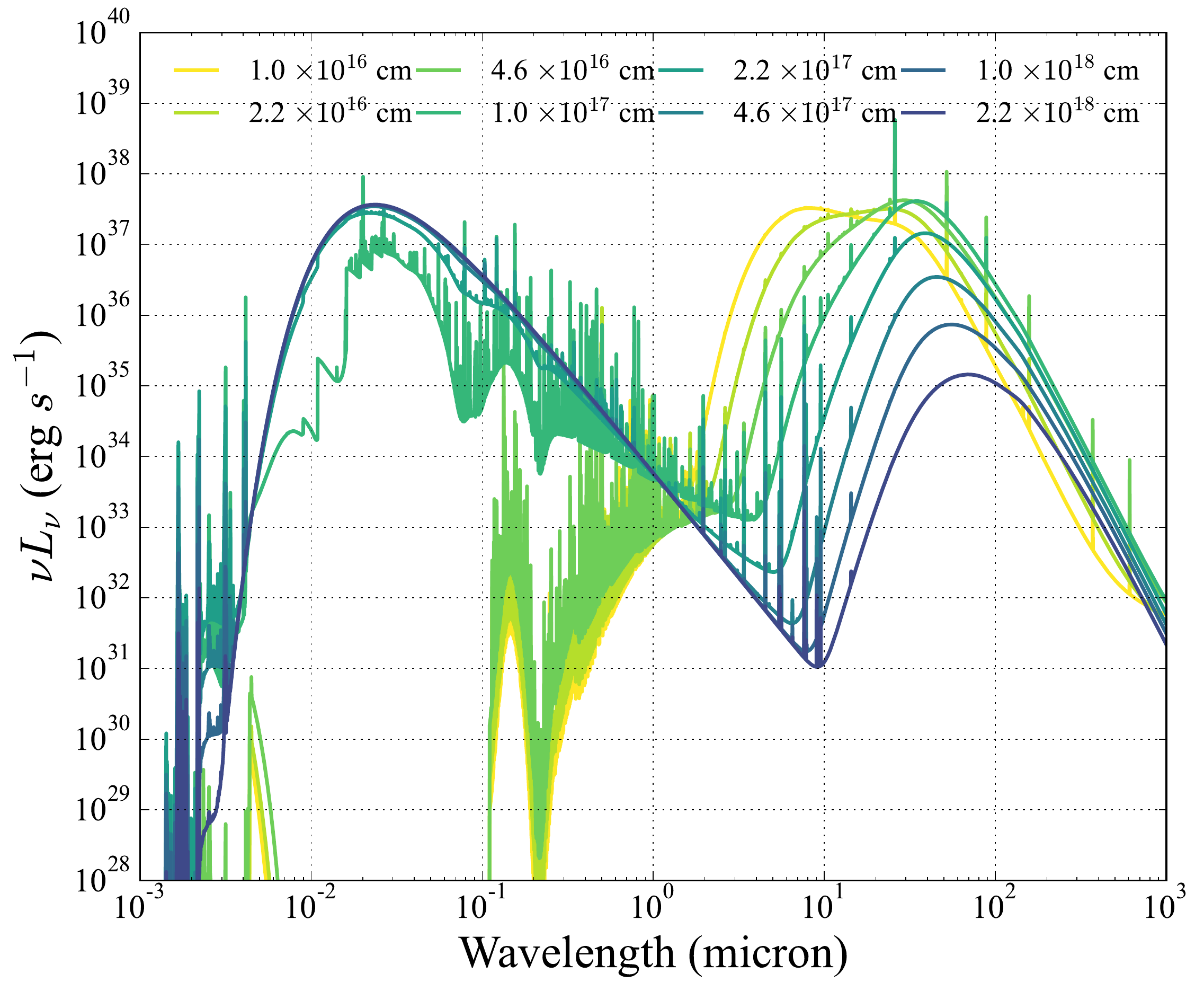}
        \includegraphics[width=1\columnwidth]{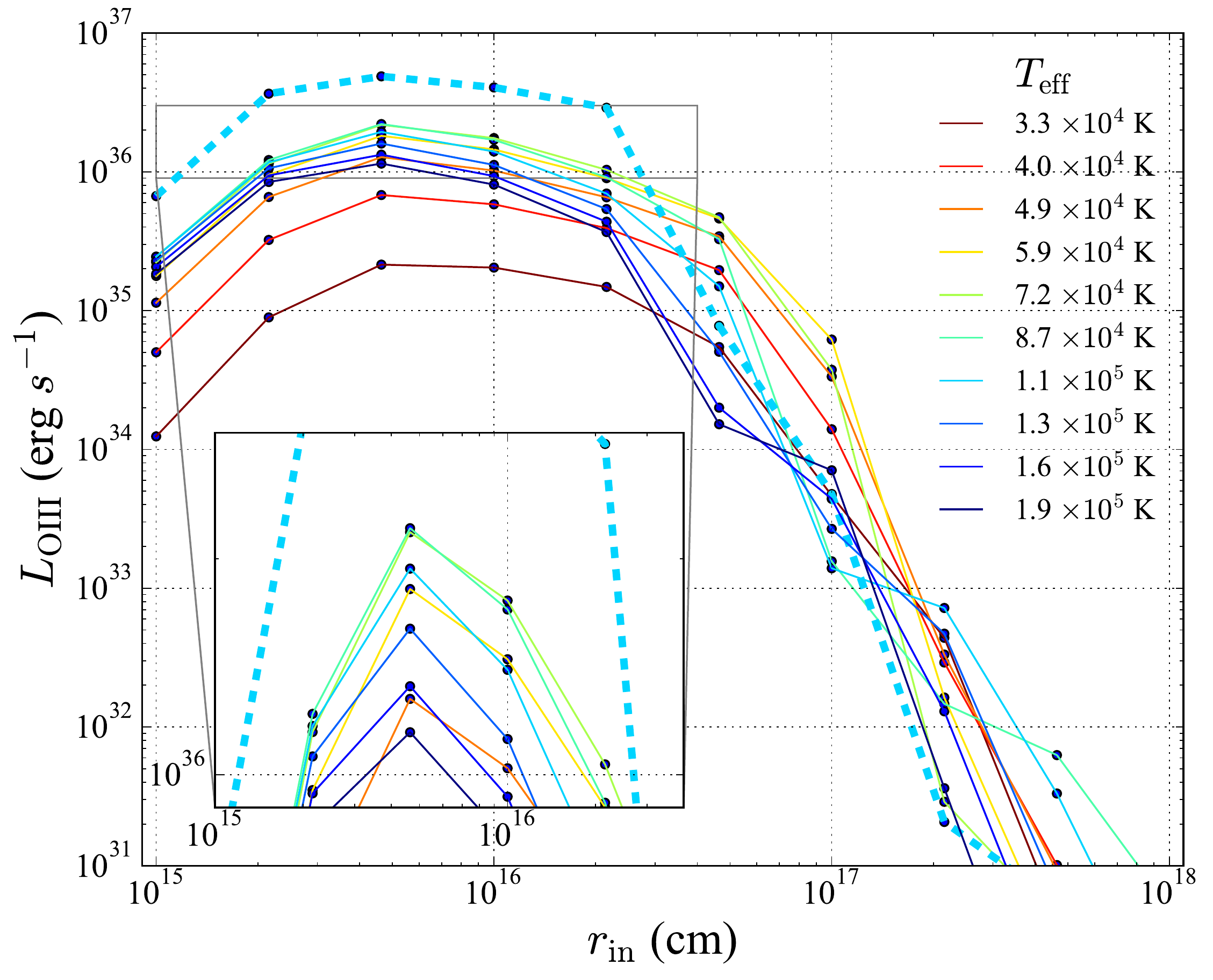}
        \caption{WD merger remnants from CLOUDY models with graphite grains and a D/G mass ratio of 10\%. The left panel shows spectra of merger remnants with a central stellar effective temperature of 150,000K, and a range of cloud inner radii from $10^{16}$cm to $2.2 \times 10^{18}$cm. The right panel shows the line luminosity of [O III] 5007 \r{A} as a function of cloud size (inner radius) from CLOUDY, assuming a D/G mass ratio of 10\%. An example model without dust at $ 1.1 \times 10^5$K is overplotted with the cyan dashed line. Different stellar effective temperatures are plotted to cover the thermal evolution of the remnant in the near-constant luminosity ($\sim 10^4 L_{\odot}$) phase (Figure \ref{fig:contaminants}).   Dust somewhat suppresses the optical line emission when the nebula is small but has little effect at larger sizes.}
        \label{fig:dust}
    \end{figure*}

    Dust may play an important role in modifying the appearance of the merger remnant due to the carbon and oxygen-rich composition of WD merger remnant ejecta.  To assess the impact of dust, we carried out the same grid of photoionization calculations as in \S \ref{sec:nodust} but for dust-to-gas (D/G) mass ratio set to 10\%.  Although the exact D/G mass ratio in merger remnant ejecta is uncertain, we assume a value that is larger than a typical planetary nebula value of $\sim$0.1\% - 1.0\% \citep{Stasinska1999} because of the higher metallicity of the merger remnant ejecta.
    
    The dust composition in the case of  C/O WD mergers depends on the relative mass in carbon and oxygen.   When the central star is a giant, the absence of UV radiation permits the formation of CO molecules in the outflow. If carbon dominates by mass,  the remaining carbon not in the form of CO will form graphite dust. On the other hand, if oxygen dominates by mass, there is remaining oxygen which forms silicate grains.  Here we assume graphite grains for concreteness, but in the next section we will separately show results for silicate grains in models of the Milky Way merger candidate IRAS 00500+6713.
    
    In the presence of dust, the short-wavelength transmitted spectrum depends on the UV optical depth to dust which is $\tau \simeq M_{\rm dust} \kappa_{\rm dust}/(4 R_{\rm out}^2)$ where $M_{\rm dust}$ is the total dust mass in the cloud of outer radius $R_{\rm out}$ and $\kappa_{\rm dust}$ is the UV optical depth per unit mass of dust.   Requiring $\tau \lesssim 1$ implies
    \begin{equation}
    R_{\rm out} \gtrsim 2 \times 10^{17} {\rm cm} \, \left(\frac{M_{\rm dust}}{\rm 0.01{\rm M_{\odot}}} \, \frac{\kappa_{\rm dust}}{10^4 \, {\rm cm^2 \, g^{-1}}}\right)^{1/2}
    \label{eq:tauUVdust}
    \end{equation}

    The left panel of Figure \ref{fig:dust} indeed shows that when the cloud radius is $\lesssim 10^{17}$cm for the C/O remnant, the grains absorb nearly all of the incident photons and re-emit them in an infrared continuum.\footnote{Recall that our nebula models assume $R_{\rm out} = 3 r_{\rm in}$ and $r_{\rm in}$ is the size-parameter shown in Figure \ref{fig:dust}.} At these stages, dust would significantly modify the appearance of the merger remnants in the CMDs in Figure \ref{fig:CMD}. For larger nebula radii, however, when the surrounding gas is photoionized and the dust is optically thin, the UV and optical continuum becomes detectable.   This likely applies during most of the time of the merger remnant evolution, and thus to most of the sources in the model CMDs in \S \ref{sec:compCMD}.

    However, when the ejecta expands to $\gtrsim 10^{17}$cm, the optical depth to dust drops below unity. In this case, the transmitted spectrum of the cloud resembles that of the dust-free scenario with the UV/optical lines becoming slightly weaker. It is important to stress that strong lines such as the [O III] 5007 line are still present even when the UV continuum is absent. This is because  the photoionization cross-section is much larger than the dust absorption cross-section so that the (carbon and oxygen) ionizing photons are initially absorbed by photoionization leading to strong optical lines even when dust ultimately absorbs and re-radiates much of the input radiation.  

    The long-wavelength infrared continuum due to dust in Figure \ref{fig:dust} is much stronger than the long-wavelength continuum due to  free-free emission in the dust-free models in Figure \ref{fig:full_spec}.   Weaker lines are no longer detectable relative to this stronger  continuum, but strong lines such as the [O IV] 25.8832 $\rm \mu m$ should still be observable. 

    The right panel of Figure \ref{fig:dust} shows the [O III] 5007 \r{A} line luminosity in our models with dust as a function of the inner radius of the ejecta for different stellar effective temperatures; we also show an example without dust for comparison (dashed cyan line).  The peak line luminosity in the case with dust is a factor of $\sim 5$ lower than without dust; nonetheless, a relatively strong line is still present even when the cloud is small and the UV dust optical depth is large (this is again due to the photoionization cross-section being much larger than the dust cross-section so that an optical-IR emitting photoionized nebula is present even when the UV dust optical depth is $\gtrsim 1$).  The peak line luminosity occurs when the cloud radius is $\sim 10^{16}$ cm both with and without dust and is predominantly set by photoionization physics.  We also note that the results in Figure \ref{fig:spec_grid} and Table \ref{tab:lines} for the most prominent lines in merger remnant PNe relative to solar composition PNe remain reasonably applicable in our models with dust, particularly the infrared lines.

\subsection{Detectability in External Galaxies}

%\EQ{NOTE:   we should also remove the MUSE lines from Figs 6 \& 7 and replace that with a discussion in this section.  i think we can keep CIV but stress that it's only detectable in teh MW not in external galaxies}

 The primary challenges in detecting the ionized nebulae predicted here in external galaxies lie in (1) the large underlying stellar continuum that dilutes the detectability of the line emission, and (2) the low surface density of merger remnants which requires surveying a relatively large area.   Here we briefly quantify these effects.

We consider an external galaxy with stellar surface mass density $\Sigma_*$ and stellar mass to light ratio in the optical of $\Upsilon$.   If observations are made with pixels (or seeing-limited resolution) of area $A_{\rm pixel}$ the total stellar luminosity per pixel is $\Sigma_* \Upsilon^{-1} A_{\rm pix}$.   For observations with spectral resolution $R = 3000 R_{3000}$ the stellar continuum background per spectral resolution element is then $\Sigma_* \Upsilon A_{\rm pix}/R$ (we scale $R$ here to the rough resolution of MUSE on the VLT, {but also discuss the case of narrow-band imaging below}).   In order to detect, say, the OIII line luminosity relative to the background stellar continuum we thus require $L_{\rm OIII} \gtrsim \Sigma_* \Upsilon A_{\rm pix}/R$, i.e.,
\begin{equation}
    \Sigma_* \lesssim \, 3 \times 10^5 \, {\rm M_\odot \, arcsec^{-2}} \left(\frac{L_{\rm OIII}}{10 L_\odot}\right) \left(\frac{A_{\rm pix}}{\rm 0.1 arcsec^2}\right)^{-1} {R_{3000} \Upsilon}
    \label{eq:Sigstar}
\end{equation}
The stellar surface density limit in equation \ref{eq:Sigstar} is comparable to the surface brightness of M31's bulge or that of M87 exterior to $\simeq 10$ kpc (note that the latter region contains $\gtrsim 3 \times 10^{11} M_\odot$; \citealt{Gebhardt2009}).

For a given stellar mass surface density the expected number of merger remnants per arcsec$^2$ $\Sigma_{\rm remnants}$ is given by
\begin{equation}
\label{eq:remnantSig}
    \Sigma_{\rm remnants} \simeq 10^{-10} \, \Sigma_* \left(\frac{ \dot{N}}{10^{-14}\,{\rm yr^{-1}}{\rm M_{\odot}}^{-1}} \frac{ t_{\rm remnant}}{\rm 10^4 \, yr}\right).
\end{equation}
where $\Sigma_*$ is in units of $M_\odot \, {\rm arcsec^{-2}}$ and where we have normalized the merger rate and remnant lifetime as in equation \ref{eq:Nactive}.   The constraint in equation \ref{eq:Sigstar} together with the expected surface density of remnants in equation \ref{eq:remnantSig} thus translates into a lower limit on the area $A$ that must be surveyed to detect of order N merger remnant nebulae:
\begin{align}
    \label{eq:Asurvey}
    A \gtrsim & 10 \, N \, {\rm arcmin^2} \, \left(\frac{ \dot{N}}{10^{-14}\,{\rm yr^{-1}}{\rm M_{\odot}}^{-1}} \frac{\rm t_{remnant}}{\rm 10^4 \, yr}\right)^{-1} \times \nonumber \\
    & \left(\frac{L_{\rm OIII}}{10 L_\odot}\right)^{-1} \left(\frac{A_{\rm pix}}{\rm 0.1 arcsec^2}\right) \Upsilon^{-1} R_{3000}^{-1}
\end{align}
Note that equation \ref{eq:Asurvey} assumes that the stellar mass satisfying equation \ref{eq:Sigstar} is $\gtrsim 10^{10} M_\odot$ so that at least one remnant is present in the galaxy of interest (eq. \ref{eq:Nactive}).   Equation \ref{eq:Asurvey} is a non-trivial constraint given that, e.g., the field of view of MUSE of 1 arcmin$^2$.  We note, however, that \citet{Sarzi2018} surveyed 33 galaxies for a total area of order $60\, {\rm arcmin^2}$ in Fornax searching for planetary nebulae, with 23 of them being ellipticals. It is thus possible that WD merger nebulae are present in this data set (eq. \ref{eq:Asurvey}).
Narrow band imagers with $R \sim 100$ can have a field of view of $\simeq$ 1 square degree (e.g., the WIYN ODI) and so are significantly more efficient at surveying a large area than IFUs.  The downside is that the lower spectral resolution restricts the observations to lower stellar surface densities (eq. \ref{eq:Sigstar}) which have less stellar mass and thus a smaller fraction of the expected WD merger remnants.

As a final point of comparison, we note that the expected total number of PNe in a galaxy would scale with the host-galaxy bolometric luminosity and the specific PN number density, which is correlated with the age and metallicity of the parent stellar population. For late-type and early-type galaxies, the total number of PNe could be $\sim 1000 \times$ and $\sim 100 \times$ higher respectively than the number of WD merger remnant nebulae present \citep{Buzzoni2006, Longobardi2013}.  The WD merger remnant nebulae are distinguishable by their absence of strong Hydrogen and Helium lines.   In addition, for galaxies with old stellar populations, the WD merger remnant nebulae may preferentially populate the more readily detectable high luminosity end of the planetary nebula luminosity function.   This is because WD merger remnants are significantly brighter than the WDs formed from solar-mass stars (Fig. \ref{fig:contaminants}); the dust-to-gas ratio in WD merger nebulae is uncertain, however, and also influences the exact optical line luminosity from the resulting nebulae (Fig. \ref{fig:dust}).

\subsection{IRAS 00500+6713}\label{sec:IRAS 00500+6713}

\renewcommand{\arraystretch}{1.2} % Default value: 1
\begin{table}
    \centering
        \begin{tabular}{|l|l|}
            \hline
            \multicolumn{2}{|c|}{Parameters of the Central Star}\\
            \hline
            $\log_{10}[L_{\rm \star}(\rm L_{\odot})]$ & $4.60 \pm 0.14$ \\
            $T_{\rm eff} (\rm K)$ & $211,000^{+40,000}_{-23,000}$\\
            Distance (\rm kpc)& $2.30\pm0.14$\\
            Shell Radius (\rm pc)& $1.1$\\
            Halo Radius (\rm pc)& $1.6$\\
            O mass fraction & $0.8 \pm 0.1$\\
            C mass fraction & $0.2 \pm 0.1$\\
            Ne mass fraction & $0.01$\\
            \hline
            \hline
            \multicolumn{2}{|c|}{Nebula Luminosity (erg/s)}\\
            \hline
            $\nu L_{\nu,W3}$ & $1.7 \times 10^{34}$\\
            $\nu L_{\nu,{\rm W4,1.1pc}}$ & $5.7 \times 10^{34}$\\
            $\nu L_{\nu,{\rm W4,1.6pc}}$ & $1.6 \times 10^{35}$\\
            \hline
            \hline
            \multicolumn{2}{|c|}{Line Luminosity (erg/s)}\\
            \hline
            $L_{\rm [Ne\,VI]}$ & $3.1 \times 10^{33}$\\
            $L_{\rm [Mg\,VII]}$ & $2.2 \times 10^{32}$\\
            \hline
        \end{tabular}
        \caption{Parameters of IRAS 00500+6713, a potential WD merger remnant, obtained from \citet{Gvaramadze2019}; distances updated based on Gaia Early Data Release 3 \citep{Bailer-Jones2021}; monochromatic luminosity  obtained from WISE images in W3 ($\lambda_{\rm ref} = 22.2 \mu \rm m$) and W4 ($\lambda_{\rm ref} = 12.1 \mu \rm m$) bands, also plotted in Figure \ref{fig:IRAS 00500+6713}; and predicted line luminosity for lines [Ne VI] 7.64$\mu \rm m$ and [Mg VII] 9.01$\mu \rm m$ within JWST MIRI's range, assuming a D/G mass ratio of 10\%. Two luminosity are provided for the W4 band depending on if halo subtraction is performed, and is discussed further in \S \ref{sec:IRAS 00500+6713}.}
        \label{tab:IRAS 00500+6713}
\end{table}

\begin{figure*}
    \centering
    \includegraphics[width=0.7\textwidth]{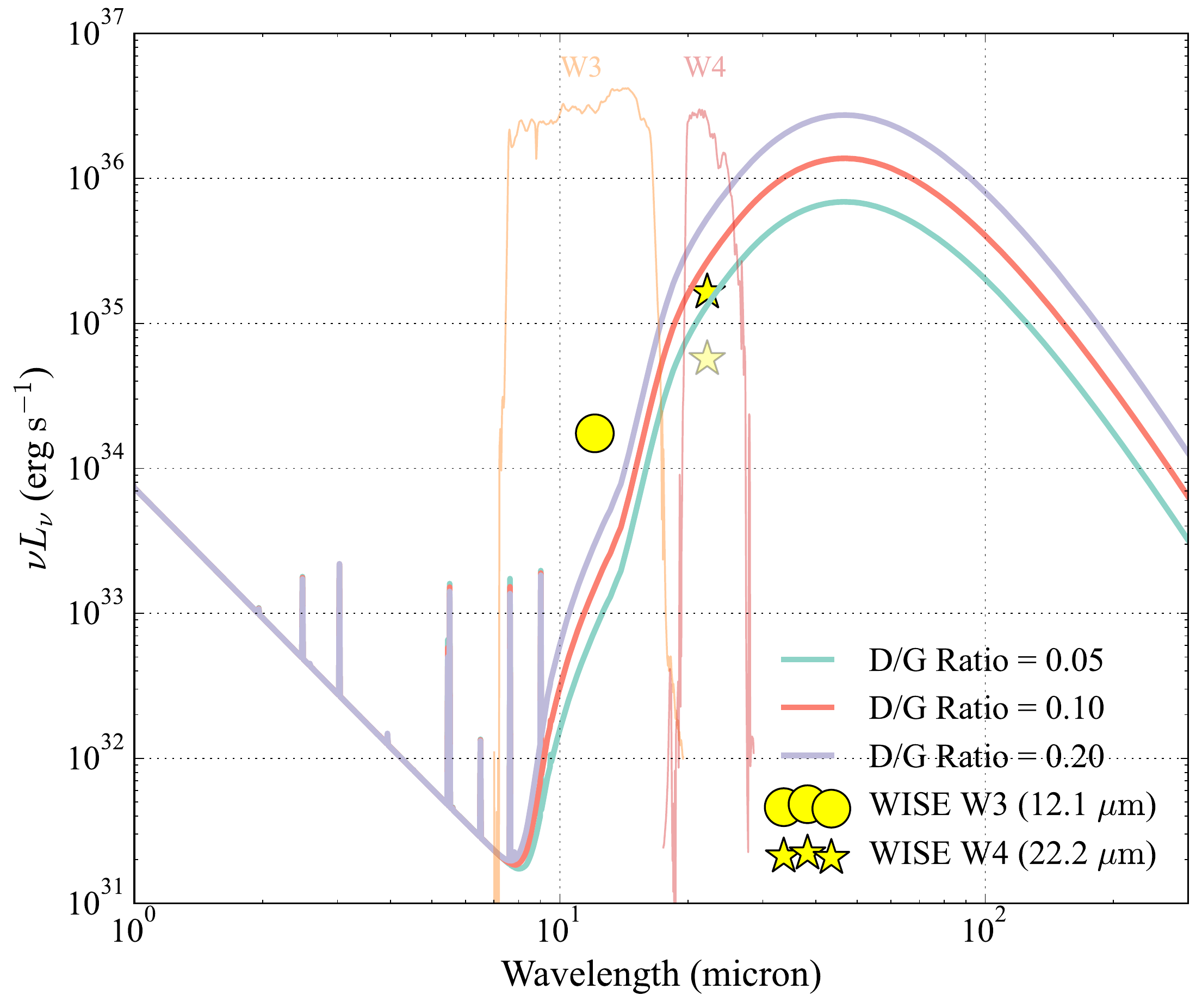}
    \caption{CLOUDY spectra for the potential WD merger remnant in the Milky Way (IRAS 00500+6713) calculated using parameters listed in Table \ref{tab:IRAS 00500+6713} with a $0.1{\rm M_{\odot}}$ ejecta mass and a range of D/G ratios. The nebula luminosity and filter response functions for WISE W3 and W4 bands are plotted in orange and red; the circular and star-shaped points represent the nebula luminosity from reduced WISE images in these bands. The lower opacity star-shaped WISE point results from a different data reduction choice for the stellar background, and is discussed more in-depth in \S \ref{sec:IRAS 00500+6713}.  Our models predict that the WISE detections are primarily dust continuum, not infrared spectral lines.}
    \label{fig:IRAS 00500+6713}
\end{figure*}

\citet{Gvaramadze2019} reported the detection of a hot luminous Galactic star surrounded by a hydrogen- and helium-free nebula (IRAS 00500+6713) that was bright in the WISE W3 ($\lambda_{\rm ref} = 22.2 \mu \rm m$) and W4 ($\lambda_{\rm ref} = 12.1 \mu \rm m$) bands.   The star was also H and He-free with a remarkable $\sim 16,000$ km s$^{-1}$ wind. Table \ref{tab:IRAS 00500+6713} summarizes some of the properties of the star and nebula reported by \citealt{Gvaramadze2019}. They interpreted this source as a pre-collapse super-Chandrasekhar WD merger remnant with the unusual wind powered by strong magnetic fields and rotation generated during the merger.

We use the properties listed in Table \ref{tab:IRAS 00500+6713} together with an assumed nebula mass of $0.1{\rm M_{\odot}}$ and several assumed D/G ratios to compute CLOUDY models of the surrounding nebula. Since the derived stellar abundance is 80\% oxygen and 20\% carbon, we assume the dust composition is primarily silicate. From X-ray spectroscopy \citet{Oskinova2020} measures a different composition in the nebula relative to \citet{Gvaramadze2019}'s optical measurements for the star (higher C and Ne fractions). The origin of this difference is unclear but introduces some systematic uncertainty in our predictions of the nebula's properties.

For our assumed parameters of IRAS 00500+6713, we find that the infrared emission is dominated by dust with only a small contribution from lines.    The properties of the infrared emission thus depend primarily on the total dust mass ($m_{\rm ej}\times$D/G ratio) and assumptions about the grain composition and size distribution.   Spectral lines between $1 \mu \rm m < \lambda < 10 \mu \rm m$ such as [Ne VI] 7.64 $\mu \rm m$ are present in all of our models, but do not contribute significantly to the WISE photometric data.    This is a property of the relatively low gas densities associated with our assumed ejecta mass of $0.1{\rm M_{\odot}}$ and the large measured nebula size.   More line features in the optical and on top of the dust continuum start to appear in models with $m_{\rm ej} \gtrsim 1.0 {\rm M_{\odot}}$ but even then dust dominates the total luminosity in the WISE bandpasses unless $m_{\rm ej} \gg 1{\rm M_{\odot}}$ or there is significant clumping. \citet{Gvaramadze2019} suggested that the observed nebula emission is dominated by [O IV] 25.89 $\mu \rm m$ and [Ne V] 14.32 $\mu \rm m$ and 24.32 $\mu \rm m$ lines. However, our models show that this is not the case unless the remnant is dust-free and has $m_{\rm ej} \gg 0.1 {\rm M_{\odot}}$, in which case it is unlikely to be the result of a double WD merger.  Follow-up mid and far-infrared spectroscopy with JWST could help clarify the nature of this unusual source. In particular, confirmation that the emission in the WISE bands is dust dominated (not line-dominated) would provide additional support for the interpretation that this object is a WD merger remnant.

\citet{Gvaramadze2019} do not report  luminosity for the nebula surrounding IRAS 00500+6713. We thus reduced the WISE data surrounding IRAS 00500+6713 by masking the foreground and background star that overlaps with the nebula and performing sky subtraction. The resulting luminosity in the WISE bands are  shown in Figure \ref{fig:IRAS 00500+6713} and Table \ref{tab:IRAS 00500+6713}.
The halo structure outside the nebula (see \citealt{Gvaramadze2019} Figure 1) in the WISE data can only be seen in W4, not in W3. Hence, subtracting the stellar background in W4 with or without the halo results in different nebula luminosity illustrated by the two different  points in Figure \ref{fig:IRAS 00500+6713}. The W4 luminosity we obtain matches our model for dust-dominated WD merger remnant spectra quite well, but the W3 luminosity is a few times higher. Overall, given the large uncertainty in dust properties including total dust mass, grain size distribution, dust clumping, etc, our model provides a reasonable explanation for the  WISE observations. 

One striking feature of the central star IRAS 00500+6713 is its remarkably high speed wind with $\dot M \simeq 3 \times 10^{-6} \,{\rm M_{\odot}} \, {\rm yr^{-1}}$ and $v_{wind} \simeq 16,000 \, {\rm km \, s^{-1}}$ \citep{Gvaramadze2019}.    This may account for one slightly puzzling feature of the WISE nebula:  the size of 1.6 pc corresponds to a wind speed of $\gtrsim 150 {\rm km \, s^{-1}}$ if the age of the system is $\lesssim 10^4$ yr, as expected for a WD merger remnant.   This is larger than the wind speeds observed for red giants and asymptotic giant branch stars \citep{Netzer1993}.  The net momentum supplied by the {current} wind of IRAS 00500+6713 over a time $t$ is $\sim 50{\rm M_{\odot}} {\rm km \, s^{-1}} \, (t/10^3 {\rm yr})$.   This is sufficient to accelerate $0.1{\rm M_{\odot}}$ of ejecta to $500 \, {\rm km \, s^{-1}}$  (if the interaction is non-radiative and conserves energy not momentum, the  speed the ejecta can be accelerated to will be even higher).   It is thus possible that the mass of the nebula around IRAS 00500+6713 was set during an earlier giant phase but that its kinematics is due to interaction with the unusually powerful current wind of the central star.  This could also account for the high expansion velocity of $\sim$ 1100 km/s measured in \citet{Fesen2023}'s [S II] optical spectra.   It is likewise worth noting that the interaction between the current high-speed wind and a previous cooler, slower, denser wind could modify the thermal and ionization state of the denser gas, an effect that is not accounted for in our photoionization models. This could explain the filamentary structure seen in the [S II] 6716 and 6731 \r{A} spectra in \citet{Fesen2023}.

\section{Summary \& Discussion}\label{sec:discussions}
It is plausible that many WD mergers do not lead to prompt thermonuclear supernovae, but instead lead to a long-lived stellar remnant.  This is particularly true for mergers of  two C/O WDs or a C/O and O/Ne/Mg WD because the temperature required to fuse Carbon is $\sim 10$ times higher than that to fuse He and so dynamical burning is  less likely during the merger.    In this paper, we have investigated the best ways of observationally detecting  surviving C/O-C/O or C/O-O/Ne/Mg merger remnants, utilizing the models of \citet{Schwab2016,Schwab2021} for the post-merger evolution of the remnant.   These models predict that the merger remnant inflates to become a giant before subsequently contracting and evolving to higher $T_{\rm eff}$ (see Figure \ref{fig:contaminants}).   The overall evolution of the merger remnant is similar to that of a $6-10 {\rm M_{\odot}}$ star as it evolves from the AGB through to the WD cooling sequence.    As a result, we highlight the value of searching for WD merger remnants in galaxies with low specific star formation rates since this implies a higher ratio of WD mergers to intermediate-mass stars.  This includes, e.g., the bulge of the MW and M31 and the outskirts of M87.

To quantify the detectability of WD merger remnants in  photometric data, we created composite stellar populations to simulate H-R diagrams  of an M87-like late-type galaxy, comparing the population of WD merger remnants to those of single and binary stellar evolution models using MIST and BPASS, respectively (see Figure \ref{fig:HR}). In principle, WD merger remnants appear to stand out as an unusually high $T_{\rm eff} \sim 10^{5.5}$ K population relative to single and binary stellar population synthesis models. However, in practice, this technique for finding WD merger remnants has several challenges. First, existing and planned UV instrumentation is not sufficiently blue to distinguish WD merger remnants from He stars formed by binary evolution (see Figure \ref{fig:CMD}). Second, and probably more severely, post-outburst classical novae undergo a very similar evolution to WD merger remnants (see Figure \ref{fig:contaminants}).   Moreover, the thermal evolution time of post-outburst novae is years-decades and they significantly outnumber WD merger remnants even in old stellar populations.   Spectroscopic information to distinguish classical novae (with hydrogen) from WD merger remnants (without hydrogen) is thus likely needed.   

Just like AGB stars generate strong dusty winds, it is likely that WD merger remnants during their giant phase generate significant outflows \citep{Schwab2016}.   We argue that the resulting photoionized nebula (analogous to a planetary nebula) is one of the most promising ways of searching for WD merger remnants.  Our CLOUDY models show that WD merger remnant nebulae have 
unique traits such as a high [O III]/$\rm H \alpha$ ratio and strong carbon and oxygen recombination lines and oxygen fine structure lines (see Figure \ref{fig:spec_grid}). The merger-remnant wind is likely even more dust-enriched than AGB winds, producing a strong mid-far-infrared continuum.   Although dust suppresses the strength of optical emission lines by a factor of $\simeq 5$ when the nebula is small (and the dust optical depth is high), the optical lines are still detectable even in this phase (Figure \ref{fig:dust}).   In addition, dust has less of an effect on the optical emission when the nebulae are larger, which is also where they spend most of their time. Photo-absorption and extinction in the UV suppress the short wavelength continuum emission when the cloud radius is $\lesssim 10^{17}$ cm, which likely precludes photometrically finding the merger remnants in these bands in the early phase of their thermal evolution.  At later times, however, the stellar continuum emission is detectable through the surrounding fully ionized nebula.

Overall, we predict that M87 should have  $\sim 100$ WD merger remnants with detectable [O III] lines (and large [O III]/$\rm H \alpha$ line ratios) in integral field unit or narrow-band imaging data.   For other old stellar populations, the number of luminous hot merger remnants detectable at a given time should scale roughly with the stellar mass of the galaxy. The primary observational challenge in detecting these ionized nebulae in external galaxies is that the background stellar continuum dilutes the line emission making it only detectable in lower stellar density regions with $\Sigma_* \lesssim 3 \times 10^5 M_\odot \, {\rm arcsec}^{-2}$ (eq. \ref{eq:Sigstar}) even with integral field unit spectrographs (with narrow-band imaging the observations are restricted to even lower stellar surface densities).   This implies that the bulge of the MW and M31 and the outskirts of nearby massive galaxies such as M87 are the most promising targets. Observations will need to survey a large area $\gtrsim 10 \, {\rm arcmin^2}$ to detect even a single merger remnant given the low surface density of such objects (eq. \ref{eq:Asurvey}.)   \citet{Sarzi2018} have already covered such an area in Fornax searching for planetary nebulae.

In the Milky Way disk, the challenge of detecting WD merger remnants is that the population of photometrically similar objects due to ongoing star formation greatly outnumbers the modest number $\sim 10$ of expected merger remnants.   Nonetheless, the latter should stand out spectroscopically by their H and He-free ionized nebulae. Indeed, \citet{Gvaramadze2019} proposed that the Galactic star IRAS 00500+6713 is a super-Chandrasekhar WD merger remnant due to its C/O-dominated composition, luminosity and effective temperature similar to the \citet{Schwab2016} merger remnant models, bright H-free WISE nebula, and remarkably high-speed stellar wind suggestive of strong magnetic fields generated during a stellar merger. Our photoionization models explain the WISE nebula surrounding IRAS 00500+6713 as dust emission; although some mid-infrared lines should be present (e.g., [Ne VI] 7.64 $\mu \rm m$) they do not contribute significantly to the WISE luminosity (see Figure \ref{fig:IRAS 00500+6713}).  Additional infrared spectroscopy with JWST would be very valuable for testing this prediction and using line diagnostics to constrain the density and electron temperature of the nebula. In particular, our fiducial $0.1{\rm M_{\odot}}$ nebula model predicts that the [Ne\,VI] line at 7.64 $\mu \rm m$ should have a luminosity $\sim 3 \times 10^{33} \, {\rm erg \, s^{-1}}$ and that the [Mg\,VII] line at 9.01 $\mu \rm m$  should have a luminosity of $\sim 2 \times 10^{32} \, {\rm erg \, s^{-1}}$. Slight variations to the dust mass within a factor of a few have little effect on the line luminosity. There are, however, uncertainties in the composition of the nebula:   \citet{Oskinova2020} infer a higher C and Ne fraction in the nebula using X-ray spectroscopy, relative to the abundances inferred from the optical stellar spectrum.   Additional mid-IR spectroscopic data would help clarify the origin of these differences and better constrain the physical conditions in the nebula.

The theoretical models and stellar population estimates in this paper could be significantly improved in future work, although some of the theoretical uncertainties are sufficiently complex that progress will likely require the detection of (or strong limits on) a sample of WD merger remnants.  In particular, our predictions of the properties of merger remnant nebulae are uncertain because of the lack of good models for (or observations of) the C/O-dominated cool winds which we believe produce the nebula (during the merger remnants' earlier giant phase).   This leads to uncertainty in the mass, dust content, and speed of the ejecta, all of which factor into the detectability of the photoionized nebula.   Separately, when estimating the number of merger remnants detectable at any time, we focused on super-Chandrasekhar mergers.   However, lower mass merger remnants have a similar evolutionary track and cover almost the same effective temperature range as their more massive counterparts (see Figure \ref{fig:contaminants}). Hence, the actual number of active remnants could be up to $\sim 10$ times higher than the $\sim 100$ we estimated for M87 if the total formation rate of long-lived merger remnants is of order the Type Ia supernova rate \citep{Maoz2012}.

\section*{Acknowledgements}
We thank Max Briel, Shany Danieli, Bruce Draine, Jan Eldridge, G\"otz Gr\"afener, Jenny Greene, Zachary Hemler, Shri Kulkarni, Norbert Langer, Ben Margalit, Ken Shen, Héloïse Stevance for helpful conversations and/or correspondence.  We are particularly grateful to Nadia Zakamska for useful conversations and comments on an initial draft of the paper.   EQ was supported in part by a Simons Investigator award from the Simons Foundation. This research benefited from interactions at workshops funded by the Gordon and Betty Moore Foundation through Grant GBMF5076.

%%%%%%%%%%%%%%%%%%%%%%%%%%%%%%%%%%%%%%%%%%%%%%%%%%
\section*{Data Availability}

The data underlying this paper will be shared on reasonable request to the corresponding author. The WISE data can be accessed via the IPAC Infrared Science Archive \url{https://irsa.ipac.caltech.edu/applications/wise/}.

%%%%%%%%%%%%%%%%%%%% REFERENCES %%%%%%%%%%%%%%%%%%

% The best way to enter references is to use BibTeX:

\bibliographystyle{mnras}
\bibliography{example} % if your bibtex file is called example.bib

% Alternatively you could enter them by hand, like this:
% This method is tedious and prone to error if you have lots of references
%\begin{thebibliography}{99}
%\bibitem[\protect\citeauthoryear{Author}{2012}]{Author2012}
%Author A.~N., 2013, Journal of Improbable Astronomy, 1, 1
%\bibitem[\protect\citeauthoryear{Others}{2013}]{Others2013}
%Others S., 2012, Journal of Interesting Stuff, 17, 198
%\end{thebibliography}

%%%%%%%%%%%%%%%%%%%%%%%%%%%%%%%%%%%%%%%%%%%%%%%%%%

%%%%%%%%%%%%%%%%% APPENDICES %%%%%%%%%%%%%%%%%%%%%

%%%%%%%%%%%%%%%%%%%%%%%%%%%%%%%%%%%%%%%%%%%%%%%%%%

% Don't change these lines
\bsp	% typesetting comment
\label{lastpage}
\end{document}